\documentclass{ws-ijmpa}
\usepackage{graphics}
\usepackage{caption}
\usepackage{subcaption}
\RequirePackage[T1]{fontenc}
\usepackage{mathtools}
\usepackage{longtable}
\usepackage{float}
\RequirePackage{graphicx}
\RequirePackage{mathptmx}      
\RequirePackage{flushend}
\RequirePackage[numbers,sort&compress]{natbib}
\RequirePackage[colorlinks,citecolor=blue,urlcolor=blue,linkcolor=blue]{hyperref}
\usepackage{longtable, subcaption, booktabs}
\usepackage{amsmath}
\usepackage{subcaption}
\usepackage[numbers]{natbib}
\usepackage{rotating}
\usepackage{lscape}
\usepackage{graphicx}
\usepackage{braket}
\usepackage{multirow}
\usepackage{amsmath}

\begin{document}

	\title{Exploring Charm Bound States: Mass Spectra and Decay Dynamics of $D$ Mesons and $Cq\bar{q}\bar{q}$ Tetraquarks”}
	\author{Chetan Lodha } 		
	\address{Department of Physics, Sardar Vallabhbhai National Institute of Technology, Surat, Gujarat-395007, India\\
		iamchetanlodha@gmail.com}
	\author{Manak Parmar } 
	\address{Department of Physics, Sardar Vallabhbhai National Institute of Technology, Surat, Gujarat-395007, India\\
		mgp.physica@gmail.com}	
	\author{Ajay Kumar Rai} 	
	\address{Department of Physics, Sardar Vallabhbhai National Institute of Technology, Surat, Gujarat-395007, India\\
		raiajayk@gmail.com}

	%
%
	%
	%

	\maketitle
	\begin{history}
		\received{Day Month Year}
		\revised{Day Month Year}
		\accepted{Day Month Year}
		\published{Day Month Year}
	\end{history}
	
	\begin{abstract}
		
		Motivated by the discovery of several charm states exhibiting tetraquark-like characteristics at BESIII and LHCb, this study investigates the spectroscopy and decay properties of $D$ mesons and tetraquark states with quark content $Cq\bar{q}\bar{q}$ within the diquark--antidiquark framework. The analysis is performed using a potential model based on the Cornell potential, considering color antitriplet--triplet configurations. Mass spectra are computed for both mesons and tetraquarks, while their decay behaviors are examined using the factorization approach for $D$ mesons and Fierz rearrangement for tetraquark decays. Theoretical results are compared with experimentally observed resonances to improve our understanding of charm quark bound systems.
		
	\end{abstract}
	
	
	\section{Introduction}
	Over the past few decades, numerous discoveries have challenged the conventional picture of hadrons, traditionally classified as either baryons or mesons \cite{Gell-Mann:1964ewy}. These findings have led to the theoretical consideration of exotic states such as tetraquarks, pentaquarks, hybrid mesons, hadronic molecules, and more \cite{Berwein:2024ztx}. Thanks to significant advancements in experimental facilities, a growing number of such bound states are being observed, greatly enriching our understanding of hadronic dynamics. The theoretical framework for multiquark states was first conceived in the 1970s and has since undergone substantial development. However, the first experimental evidence for such exotic matter came in 2003 at the Belle experiment, with the observation of a narrow resonance at 3872 MeV in the decay channel $B^{\pm}\rightarrow K^{\pm}J/\psi\pi^{+}\pi^{-}$~\cite{Belle:2003nnu}. This state, now known as $X(3872)$, displayed properties inconsistent with traditional meson interpretations. Subsequently, many other unconventional resonances were discovered, including $Z(4430)$~\cite{Belle:2014nuw}, $Y(4660)$~\cite{Belle:2007umv} by Belle, $Y(4140)$ at Fermilab~\cite{CDF:2009jgo}, $X(5568)$ by the DØ experiment~\cite{D0:2016mwd}, $Z_{c}(3900)$ by BESIII~\cite{BESIII:2013ris}, and $X(6900)$ by LHCb~\cite{LHCb:2020bwg}. Most of these observed tetraquark candidates include at least one heavy quark; however, several theoretical studies in recent years have also predicted the existence of all-light tetraquark states.
	
	Since their discovery at SLAC in 1976, \( D \) mesons have played a pivotal role in understanding heavy-light systems~\cite{Goldhaber:1976xn}. LHCb has reported both CP violation and mixing in the \( D^0\text{--}\bar D^0 \) system~\cite{LHCb:2019hro,LHCb:2021ykz}, phenomena that are highly suppressed in the Standard Model (SM) due to GIM cancellations and up-quark degeneracy~\cite{Bigi:2000wn}. These effects provide sensitive tests for new physics. Experiments such as LHCb, BESIII, Belle II, and PANDA are actively probing charm decays, lifetimes, and angular distributions~\cite{Owen:2024qij,Belle-II:2018jsg,Gotzen:2024agc}. Precise measurements of mixing parameters \( x \) and \( y \) from HFLAV~\cite{HeavyFlavorAveragingGroupHFLAV:2024ctg} and time-dependent studies from LHCb~\cite{LHCb:2022cak} show no significant CPV, consistent with SM expectations. Direct CP violation has also been observed at LHCb~\cite{LHCb:2019hro}, with asymmetries in \( D^0\to K^-K^+ \) and \( \pi^-\pi^+ \) exceeding naive SM predictions~\cite{Belle:2024vho}. 
	
	Theoretically, \( D \) meson decays and mixing have been studied via HQET, OPE, and lattice QCD~\cite{Bigi:2000wn,Buccella:1994nf,Falk:2001hx,Grossman:2006jg}. CP violation in charm is explored in SM and BSM frameworks such as SUSY and extended Higgs models~\cite{Golowich:2007ka,Isidori:2011qw}. The spectroscopy of excited \( D \) and \( D_s \) mesons, including the enigmatic \( D_{s0}^*(2317) \) and \( D_{s1}(2460) \), challenges conventional quark-model assignments and has prompted extensive studies. These include coupled-channel dynamics and chiral effective approaches, as well as refined quark potential and Regge-trajectory models~\cite{Godfrey:1985xj,Moir:2016srx,Patel:2022hhl,Oudichhya:2024ikt,Devlani:2013kta,Kher:2017wsq,Ni:2021pce}. Analyses of \( D \to VV \) decays further probe hadronic dynamics and QCD factorization~\cite{Cheng:2010vk,Korner:1992wi}. These theoretical efforts aim to clarify long-distance contributions and test charm mesons as probes for new physics.
	
	Charm meson spectroscopy has uncovered states beyond quark-model predictions. In the charm–strange sector, the narrow \( D_{s0}^*(2317)^\pm \) and \( D_{s1}(2460)^\pm \) lie well below potential model expectations and decay via isospin-violating modes~\cite{BaBar:2003oey,CLEO:2003ggt,Belle:2003guh}. With \( J^P=0^+ \) and \( 1^+ \), respectively, they exhibit masses around 2318 MeV and 2460 MeV and widths below 4 MeV, suggesting exotic interpretations such as DK molecules or tetraquarks~\cite{ParticleDataGroup:2024cfk,vanBeveren:2006st}. The \( D_{s1}(2700)^\pm \), consistent with a \( 2^3S_1 \) state, and two overlapping states near 2860 MeV—\( D_{s1}^*(2860)^- \) and \( D_{s3}^*(2860)^- \)—fit the D-wave excitation scheme~\cite{BaBar:2006gme,LHCb:2014ott,LHCb:2016lxy}. In the non-strange sector, the broad \( D_0^*(2400) \) (\( J^P=0^+ \)) and \( D_1(2430) \) are anomalous, while \( D_1(2420) \) and \( D_2^*(2460) \) are narrow and model-consistent~\cite{FOCUS:2003gru}.
	
	LHCb has identified open-charm tetraquark candidates like \( X_0(2900)^0 \) and \( X_1(2900)^0 \) in \( B^+\to D^+D^-K^+ \), with \( J^P=0^+,1^- \), decaying to \( \bar{D}K \)~\cite{LHCb:2020bls}. Later, \( T_{cs0}(2900)^{++,0} \) states were observed in \( B\to DD_s\pi \), forming an isospin triplet and confirming their exotic nature~\cite{Wang:2023hpp}. A significant breakthrough is the discovery of the doubly charmed \( T_{cc}^+ \), a near-threshold, weakly bound \( D^0D^{*+} \) molecule with \( J^P=1^+ \) and width in the keV range~\cite{LHCb:2021vvq}. Theoretical studies use diquark–antidiquark models, QCD sum rules, and lattice QCD to predict tetraquark spectra~\cite{Ali:2019roi,Chen:2020eyu,Wang:2020xyc,Agaev:2020mqq,Padmanath:2015era}. Hadronic molecule interpretations and coupled-channel analyses explain binding and decay properties via meson loop effects~\cite{Meng:2021jnw,Fleming:2021wmk,Ortega:2016mms}. These diverse approaches highlight the role of color correlations, threshold dynamics, and multiquark interactions in shaping the charm sector~\cite{Esposito:2016noz,Liu:2019zoy}.
	
	The present paper is organized as follows: Section I provides a brief introduction to the motivation and context of the study. Section II outlines the theoretical framework used for determining the mass spectra of mesons, diquarks, and tetraquarks. Section III discusses and analyzes the mass spectroscopy of diquarks, tetraquarks, and the $D$ meson in detail. Section IV focuses on the decay properties of mesons and tetraquarks, including both weak and strong decay channels. In Section V, Regge trajectories are constructed and examined. The results and their implications are presented in Section VI, followed by the conclusion in Section VII.
	\section{Theoretical Framework}
	\label{sec:1}
	This study builds upon our earlier investigations of tetraquark systems by introducing a charm quark in place of one of the light quarks in a previously analyzed light–light tetraquark configuration~\cite{Lodha:2025ffp,Lodha:2024}. Anchored in the theoretical framework of inter-quark dynamics and their phenomenological implications, this work follows the analytical and computational strategies developed in~\cite{Lodha:2024bwn,Lodha:2024qby}. Specifically, we consider a diquark–antidiquark configuration featuring a heavy–light pairing, which is modeled using both semi-relativistic and non-relativistic approaches to effectively capture the internal structure and mass spectrum of the tetraquark system.
	
	To evaluate the mass and binding energies of the resulting states, we employ a modified time-independent radial Schrödinger equation. The inherent two-body problem is reformulated in the center-of-mass (COM) frame, thereby reducing it to an effective central potential problem, as systematically discussed in~\cite{Lucha:1995zv}. The radial and angular components of the wavefunction are separated using spherical harmonics to solve the Schrödinger equation in spherical coordinates, enabling precise determination of eigenstates under central potential models.
	
	The quark–quark and quark–antiquark interactions within the tetraquark are modeled using phenomenological potentials inspired by QCD, including confinement and Coulomb-like terms. The semi-relativistic and non-relativistic Hamiltonians incorporate relativistic corrections for the motion of constituent quarks, crucial in systems where light and charm quarks coexist. This framework is consistent with models explored in~\cite{Esposito:2016noz}, and the full form of the Hamiltonian in the COM frame is expressed as:
	
	\begin{equation}
		H_{SR} = \sum_{i=1}^2 \sqrt{M^{2}_{i} + p^{2}_{i}+} + V^{(0)} (r) + V_{SD}(r),
	\end{equation}

	\begin{equation}
		H_{NR} = \sum_{i=1}^2 (\frac{p_{i}^{2}}{2M_{i}} + M_{i}) + V^{(0)} (r) + V_{SD}(r).
	\end{equation}

	where \( M_i \), \( p_i \), \( V^0(r) \), and \( V_{SD}(r) \) denote the constituent quark masses, the relative momentum of the bound state system, the central interaction potential, and the spin-dependent potential, respectively. The kinetic energy term in the Hamiltonian is expanded up to \(\mathcal{O}(p^6)\), capturing relativistic corrections beyond the leading-order approximation. This extended expansion accounts for higher-order momentum contributions, enabling a more precise theoretical treatment—particularly essential for systems involving charm quarks, where non-negligible relativistic effects and intricate quark-gluon dynamics play a crucial role.
	
	By incorporating these corrections, the model improves upon conventional non-relativistic approaches, offering a more reliable prediction of mass spectra and decay properties. Such precision is vital in exploring the internal structure and binding characteristics of exotic hadrons, especially tetraquark configurations where complex interplay among constituent quarks occurs. A comprehensive discussion on the significance of the higher-order momentum terms is presented in~\cite{Patel:2022hhl}. The explicit form of the kinetic energy expansion used in our analysis is:
	
	\begin{equation}
		K.E. = \sum_{i=1}^{2} \left[ \frac{p^{2}}{2M_i} - \frac{p^{4}}{8M_i^{3}} + \frac{p^{6}}{16M_i^{5}} \right].
	\end{equation}
	
	In modeling the interquark forces within the tetraquark system, we adopt the Cornell potential as the leading-order interaction, a choice well-supported by its success in conventional meson spectroscopy. This potential combines a short-range Coulomb-like term, arising from one-gluon exchange and interpreted as a Lorentz vector interaction, with a long-range linear confinement term, often treated as a Lorentz scalar. The Coulombic term dominates at small distances, reflecting the attractive force due to color charge interactions among quarks. Conversely, the linear term ensures quark confinement at large separations, representing the non-perturbative QCD regime wherein the potential energy increases linearly with distance.
	
	This dual-component Cornell potential serves as a foundational framework for understanding both conventional hadrons and exotic multiquark states. It effectively captures the essence of quark confinement and provides a practical tool for analytical and numerical evaluations of bound state energies. The explicit form of the zeroth-order Cornell-like potential is given by:
	
	\begin{equation}
		V^0(r) = -\frac{4}{3} \frac{\alpha_s}{r} + \sigma r + C,
	\end{equation}
	
	where \( \alpha_s \) is the strong coupling constant, \( \sigma \) denotes the string tension characterizing confinement, and \( C \) is a phenomenological constant adjusted to fit the ground state mass. This potential, when supplemented with spin-dependent terms \( V_{SD}(r) \), allows for a comprehensive description of fine and hyperfine mass splittings, further enhancing the predictive capability of the model for multiquark configurations such as singly charmed tetraquarks (\( Cqqq \)).
	\begin{equation}
		V^{(0)}_{C+L}(r) = \frac{k_{s}\alpha_{s}}{r} + br + V_{0},
	\end{equation}
	
	where $\alpha_{s}$, $k_{s}$, $b$ and $V_{0}$ are the QCD running coupling constant, color factor, string tension and constant, respectively. Different hadronic color configurations correspond to distinct color factors that govern their interactions. Singlet states, being color-neutral, have a color factor \( k_s = -\frac{4}{3} \), while triplet-antitriplet states carry \( k_s = -\frac{2}{3} \). Sextet states, with a more complex color structure, exhibit \( k_s = \frac{10}{3} \). These factors determine the strength and nature of color interactions, offering crucial insights into their quantum properties within QCD. A detailed discussion on color factors can be found in \cite{Muta:2010xua}. Inspired by \cite{Koma:2006si}, relativistic mass corrections modify the central potential as:\begin{equation}
		V^{(0)}(r) = V^{0}_{C+L}(r) + V^{1}(r) \biggl(\frac{1}{m_{1}} + \frac{1}{m_{2}} \biggl)
	\end{equation}
	
	where $m_{1}$ and $m_{2}$ denote the constituent masses of constituent particles in the bound state. Given the limited exploration of the non-perturbative form of the relativistic mass correction term, the leading-order perturbative approach is adopted. Specifically, the first-order approximation within perturbation theory is employed. The expression for this leading-order correction is given by:
	\begin{equation}
		V^{1}(r) = - \frac{C_{F}C_{A}}{4} \frac{\alpha_{s}^{2}}{r^{2}},
	\end{equation}
	
	where the Casimir charges $C_{F}$ and $C_{A}$ correspond to the fundamental and adjoint representations, with values of $\frac{4}{3}$ and $3$, respectively \cite{Koma:2006si}. 
	
	In addition, the inclusion of spin-dependent terms within the model has been addressed using a perturbative approach, as outlined in~\cite{Lucha:1991vn}. This methodology significantly enhances our understanding of the energy splittings observed between orbital and radial excitations across various hadronic states. By incorporating spin-dependent interactions, the model achieves a more refined and precise description of the mass spectra and internal dynamics of these systems. This expanded framework not only yields improved theoretical predictions but also facilitates deeper insights into the underlying mechanisms of quark confinement and strong interaction dynamics as governed by quantum chromodynamics (QCD).
	
	The spin-dependent interactions employed in this study are derived from the Breit-Fermi Hamiltonian for one-gluon exchange, as detailed in~\cite{Lucha:1991vn,Voloshin:2007dx}. These interactions are treated as perturbative energy corrections using first-order perturbation theory, allowing for systematic incorporation into the spectrum calculations. The relevant spin-dependent terms include the spin-spin interaction potential \( V_{SS}(r) \), which governs hyperfine splittings; the tensor interaction potential \( V_T(r) \), which affects the angular distribution of internal spin configurations; and the spin-orbit interaction potential \( V_{LS}(r) \), which contributes to the fine splitting of states based on total angular momentum.
	\begin{equation}
		V_{SD}(r)  = V_{T}(r) + V_{LS}(r) + V_{SS}(r). 
	\end{equation}
	\begin{subequations}
		\begin{equation}
			= \biggl( - \frac{k_{s}\alpha_{s}}{4} \frac{12\pi}{M_{\mathcal{D}}M_{\bar{\mathcal{D}}}}\frac{1}{r^{3}}\biggl) \; \biggl(-\frac{1}{3}(S_{1}\cdotp S_{2}) + \frac{(S_{1} \cdotp r) {(S_{2}\cdotp r)}}{r^{2}}\biggl) 
			\label{eqvt}
		\end{equation}
		\begin{equation}
			+ \biggl(-\frac{3\pi k_{s}\alpha_{s}}{2M_{\mathcal{D}}M_{\bar{\mathcal{D}}}}\frac{1}{r^{3}}  -  \frac{b}{2M_{\mathcal{D}}M_{\bar{\mathcal{D}}}}\frac{1}{r}   \biggl)(L\cdotp S) 
			\label{eqls}
		\end{equation}
		\begin{equation}
			+ \biggl(- \frac{k_{s}\alpha_{s}}{3} \frac{8\pi}{M_{\mathcal{D}}M_{\bar{\mathcal{D}}}} \frac{\sigma}{\sqrt{\pi}}^{3} exp^{-\sigma^{2}r^{2}}\biggl) (S_{1}\cdotp S_{2})
			\label{eqss}
		\end{equation} 
	\end{subequations}
	
	Here, the masses of the diquark and antidiquark are denoted by \( M_{\mathcal{D}} \) and \( M_{\bar{\mathcal{D}}} \), respectively. The values of \( (S_{1} \cdot S_{2}) \) are determined by solving the diagonal matrix elements for spin-\(\frac{1}{2}\) and spin-1 particles, as discussed in~\cite{Debastiani:2017msn}. Additionally, the parameter \( \sigma \) is introduced as a smoothed replacement for the Dirac delta function, ensuring a more tractable mathematical formulation while preserving the essential physical characteristics of the interaction.
	
	Equation~\ref{eqvt} in \( V_{SD}(r) \) represents the tensor interaction potential, while equation~\ref{eqls} corresponds to the spin-orbit interaction potential, and equation~\ref{eqss} describes the spin-spin interaction potential. These spin-dependent interactions have been extensively discussed in our previous work~\cite{Lodha:2024qby}, with additional insights provided in studies such as~\cite{Tiwari:2021iqu, Debastiani:2017msn, Lundhammar:2020xvw}. The spin-orbit potential \( V_{LS} \) and the tensor potential \( V_T \) contribute to the fine structure of the states, introducing small corrections that affect their energy levels. In contrast, the spin-spin interaction potential \( V_{SS} \) is responsible for the hyperfine splitting, distinguishing states with identical spatial configurations but different total spins.
	
	These spin-dependent interactions are typically treated within first-order perturbation theory, where their matrix elements are included as energy corrections to the total potential. Each of these interaction potentials is derived in terms of the static quark potential, \( V(r) \), which describes the quark-quark interaction as a function of their separation distance. To improve the accuracy of tetraquark spectroscopy, the spin-spin interaction can be directly included in the zeroth-order potential, offering a reasonable approximation of the mass spectra. This approach ensures that the significant contributions of spin-dependent forces, responsible for both fine and hyperfine structures, are adequately captured in the description of tetraquark states.
	
	\section{Spectroscopy}
	\subsection{Meson Spectra}
	
	We begin by computing the mass spectra of \( D \) meson, which serve as a reference for determining the fitting parameters used in the analysis of diquarks and tetraquarks. The optimized fitting parameters $b,\alpha_{s},M_{c}$ and $M_{q}$ adopted in this study are $0.1075 \pm 0.0025$ GeV$^{2}$, $0.675\pm0.01$, $0.37\pm0.01$ GeV and $1.45\pm0.025$ GeV for non-relativistic formalism and $0.107\pm0.001$ GeV$^{2}$, $0.67\pm0.01$, $0.37\pm0.01$  GeV and $1.475\pm0.025$ GeV for semi-relativistic formalism. Within the framework of SU(3) color symmetry, only color-neutral quark combinations are allowed to form a singlet state. With a color factor of \( k_s = -\frac{4}{3} \), mesons adhere to the representation \( \bar{\textbf{3}} \otimes {\textbf{3}} = \textbf{1} \oplus \textbf{8} \) \cite{Debastiani:2017msn}. The masses of the individual states within the \( [C\bar{q}] \) meson system are given by:
	
	\begin{equation}
		M_{(C\bar{q})} =  M_{C} + M_{\bar{q}} + E_{(C\bar{q})} + \braket{V^{1}(r)}.
	\end{equation}

	Here, \( M_C \) and \( M_q \) represent the constituent masses of the charm quark/antiquark and the up/down quark/antiquark, respectively. The final mass of the tetraquark state, as calculated in this study, incorporates both spin-dependent interaction contributions and relativistic corrections to the kinetic energy. These corrections are essential for accurately capturing the internal dynamics and binding mechanisms of systems involving heavy and light quarks. The mass spectra obtained from the present model are systematically compared with results from various other theoretical studies, as summarized in Table~\ref{mass_meson}, exhibiting a strong level of consistency and validating the reliability of our approach.
	
	\begin{center}
		
		\begin{longtable}{cc|cc|ccccccc}
			\label{mass_meson}\\
			\caption{Mass spectra and comparison with various theoretical models for $D$ meson in MeV} \\
			\hline

			State & $J^{PC}$& \multicolumn{1}{c}{Mass$_{NR}$} & \multicolumn{1}{c|}{Mass$_{SR}$} &\cite{Patel:2022hhl} &\cite{Devlani:2013kta}&\cite{Kher:2017wsq}&\cite{Oudichhya:2024ikt} &\cite{Godfrey:1985xj}& \cite{Ni:2021pce}\\
			\hline
			\endfirsthead
			
			\hline
			State & $J^{PC}$& \multicolumn{1}{c}{Mass$_{NR}$} & \multicolumn{1}{c|}{Mass$_{SR}$} &\cite{Patel:2022hhl} &\cite{Devlani:2013kta}&\cite{Kher:2017wsq}&\cite{Oudichhya:2024ikt} &\cite{Godfrey:1985xj}& \cite{Ni:2021pce}\\
			\hline
			\endhead
			
			\hline
			\multicolumn{6}{r}{Continued on next page} \\
			\endfoot
			
			\endlastfoot
			
			$1 ^{1}S_{0}$ &{$0^{-}$} &$1867.49\pm0.33$&$1863.67\pm0.25$ &1889

			&1865&1884&1864	&1877	&1865\\
			$2 ^{1}S_{0}$ &{$0^{-}$} &$2541.56\pm0.37$&$2532.01\pm0.22$&2601

			&2598 &2582 &2549	&2581	&2547&\\
			$3 ^{1}S_{0}$ &{$0^{-}$} &$3010.06\pm0.41$&$2937.49\pm0.18$&3108
			
			&3087 &3186 &3049	&3068	&3029&\\
			$4 ^{1}S_{0}$ &{$0^{-}$} &$3403.72\pm0.46$&$3196.06\pm0.28$&3506
			&3498 &3746 &3540	&3468	&3498&\\
			
			&&&&&\\
			
			$1 ^{3}S_{1}$ &$1^{-}$ &$2006.72\pm0.46$ &$2005.65\pm0.47$ &2007

			&2018&2010&2007	&2041	&2008	&\\
			$2 ^{3}S_{1}$ &$1^{-}$ &$2617.95\pm0.48$ &	$2611.43\pm0.41$&2631
			&2639&2655	&2627.0	&2643	&2636
			
			&\\
			$3 ^{3}S_{1}$ &$1^{-}$ &$3068.16\pm0.54$ &$3013.26\pm0.40$ &3122
			&3110 &3239 &3126	&3110	&3093	&\\
			$4 ^{3}S_{1}$ &$1^{-}$ &$3452.04\pm0.61$ & $3284.65\pm0.55$&3514
			&3514 &3789 &3556	&3497	&3514	&\\
			
			&&&&&\\
			
			$1 ^{1}P_{1}$ &$1^{+}$ &$2411.54\pm0.46$ &$2421.59\pm0.42$&2448
			&2434 &2425 &2359	&2456	&2453&\\
			$2 ^{1}P_{1}$ &$1^{+}$ &$2887.00\pm0.50$ & $2863.83\pm0.39$&2978
			&2940 &3016 &2932	&2924	&2936	&\\
			$3 ^{1}P_{1}$ &$1^{+}$ &$3286.15\pm0.57$ & $3178.97\pm0.44$&3397
			&3567 &- &3410	&3328	&-	&\\
			$4 ^{1}P_{1}$ &$1^{+}$ &$3641.62\pm0.64$ & $3365.82\pm0.76$&3737
			&- &- &3829	&3681	&-&\\
			
			&&&&&\\
			
			$1 ^{3}P_{0}$ &$0^{+}$ &$2340.18\pm0.48$ &$2354.08\pm0.44$&2382
			&2352&2357&-&2399&2313&\\
			$2 ^{3}P_{0}$ &$0^{+}$ &$2820.15\pm0.51$ &	$2804.87\pm0.41$&2937
			&2868&2976&-&2931&2849 \\
			$3 ^{3}P_{0}$ &$0^{+}$ &$3221.13\pm0.57$ &	$3129.33\pm0.45$&3367
			&-&3536&-&3343&\\
			$4 ^{3}P_{0}$ &$0^{+}$ &$3577.50\pm0.64$ &	$3329.64\pm0.72$&3713
			&-&-&-&3697&\\
			
			&&&&&\\
			
			$1 ^{3}P_{1}$ &$1^{+}$ &$2442.00\pm0.45$ &$2456.33\pm0.41$&2450
			&2454&2447&-&2467&2453\\
			$2 ^{3}P_{1}$ &$1^{+}$ &$2917.89\pm0.50$ &	$2902.33\pm0.39$&2979
			&2951 &3034 &- &2961&2936\\
			$3 ^{3}P_{1}$ &$1^{+}$ &$3317.10\pm0.57$ &	$3224.67\pm0.43$&3398
			&- &3582&-&3328\\
			$4 ^{3}P_{1}$ &$1^{+}$ &$3672.46\pm0.64$ &	$3423.75\pm0.70$&3738
			&&&&4709\\
			
			&&&&&\\
			
			$1 ^{3}P_{2}$ &$2^{+}$ &$2491.91\pm0.45$ &$2508.53\pm0.40$&2462
			&2473&2461&2405	&2502	&2475	&\\
			$2 ^{3}P_{2}$ &$2^{+}$ &$2971.84\pm0.49$ &$2957.59\pm0.38$&2985
			
			&2971 &3039 &	3012	&2957	&2955	
			&\\
			$3 ^{3}P_{2}$ &$2^{+}$ &$3373.47\pm0.56$ &$3281.85\pm0.41$&3402
			&-&3584	&3516	&3353	&-	&\\
			$4 ^{3}P_{2}$ &$2^{+}$ &$3730.48\pm0.63$ &	$3482.25\pm0.69$&3741
			&- &- &3956	&3701 &-&\\
			
			&&&&&\\

			$1 ^{1}D_{2}$&$2^{-}$  &$2721.85\pm0.47$ &$2726.65\pm0.40$&2754
			&2722&2754&2767	&2816	&2827	
			&\\
			$2 ^{1}D_{2}$ &$2^{-}$ &$3137.95\pm0.53$ &$3089.99\pm0.39$&3224
			&3169&3318&3259	&3212	&3221
			&\\
			$3 ^{1}D_{2}$ &$2^{-}$ &$3504.75\pm0.60$ &	$3440.62\pm0.53$&3600
			&-&3854&3686	&3566	&-	
			&\\
			$4 ^{1}D_{2}$ &$2^{-}$ &$3839.10\pm0.68$ &	$3458.47\pm0.99$&3906
			&-&-&4069	&-	&-	
			&\\
			
			&&&&&\\

			$1 ^{3}D_{1}$ &$1^{-}$ &$2717.81\pm0.42$ &$2722.63\pm0.41$ &2751
			&2803&2755&-&2833&2754
			&\\
			$2 ^{3}D_{1}$ &$1^{-}$ &$3132.10\pm0.54$ &	$3085.68\pm0.40$&3223
			&3233 &3215 &-&3231 &3143 	&\\
			$3 ^{3}D_{1}$ &$1^{-}$ &$3497.85\pm0.61$ &	$3337.94\pm0.53$&3600
			&- &3850 &-&3579&
			&\\
			$4 ^{3}D_{1}$ &$1^{-}$ &$3831.52\pm0.68$ &	$3459.67\pm0.98$&3906
			&- &- &-&-&-
			&\\
			
			&&&&&\\
			
			$1 ^{3}D_{2}$ &$2^{-}$ &$2727.02\pm0.47$ &$2733.44\pm0.39$&2782
			&2829&2873 &- &2845&2827\\
			$2 ^{3}D_{2}$ &$2^{-}$ &$3145.99\pm0.53$ &	$3100.75\pm0.39$&3246
			&3256&3341&-&3248&3221\\
			$3 ^{3}D_{2}$ &$2^{-}$ &$3514.74\pm0.60$ &	$3355.75\pm0.52$&3617
			&-&3873 &- &3600\\
			$4 ^{3}D_{2}$ &$2^{-}$ &$3850.54\pm0.68$ &	$3479.43\pm0.97$&3920 &-&-&-&-\\
			
			&&&&&\\
			
			$1 ^{3}D_{3}$ &$3^{-}$  &$2715.60\pm0.47$ &$2724.85\pm0.39$&2807
			&2741&2788&2745	&2833	&2782
			&\\
			$2 ^{3}D_{3}$ &$3^{-}$  &$3142.38\pm0.51$ &$3099.38\pm0.37$&3265
			&3187&3355 &3351	&3226	&3202\\
			$3 ^{3}D_{3}$ &$3^{-}$  &$3516.10\pm0.59$ &$3358.97\pm0.50$&3631
			&- &3885 &3835	&3579	&-\\
			$4 ^{3}D_{3}$ &$3^{-}$  &$3855.43\pm0.66$ &$3485.97\pm0.95$&3933
			&-&-&4277	&-	&-\\
			
			&&&&&\\
			
			$1 ^{1}F_{3}$ &$3^{+}$  &$2977.79\pm0.50$ &$2962.94\pm0.39$&3051
			
			&&&&3108&3022\\
			$2 ^{1}F_{3}$ &$3^{+}$  &$3357.20\pm0.57$ &$3261.95\pm0.43$&3472
			
			&&&&3461\\
			$3 ^{1}F_{3}$ &$3^{+}$  &$3700.84\pm0.64$&$3444.60\pm0.72$&3807
			&&\\
			
			&&&&&\\
			
			$1 ^{3}F_{2}$ &$2^{+}$  &$2982.70\pm0.49$	&$2967.18\pm0.40$&3080
			&&&&3132&3096\\
			$2 ^{3}F_{2}$ &$2^{+}$  &$3360.41\pm0.57$	&$3264.90\pm0.44$&3494
			&&&&3490\\
			$3 ^{3}F_{2}$ &$2^{+}$  &$3702.89\pm0.65$	&$3447.08\pm0.72$&3825
			&&\\
			
			&&&&&\\
			
			$1 ^{3}F_{3}$ &$3^{+}$  &$2966.44\pm0.50$	&$2953.26\pm0.39$&3048
			&&&&3143&3129\\
			$2 ^{3}F_{3}$ &$3^{+}$  &$3349.42\pm0.56$	&$3255.89\pm0.42$&3469
			&&&&3498\\
			$3 ^{3}F_{3}$ &$3^{+}$  &$3695.60\pm0.63$	&$3441.51\pm0.70$&3805
			&&&&\\
			
			&&&&&\\
			
			$1 ^{3}F_{4}$ &$4^{+}$  &$2934.96\pm0.49$	&$2924.72\pm0.38$&3029
			
			&&&&3113&3034\\
			$2 ^{3}F_{4}$ &$4^{+}$  &$3324.74\pm0.55$	&$3234.06\pm0.40$&3454
			
			&&&&3466\\
			$3 ^{3}F_{4}$ &$4^{+}$  &$3675.95\pm0.62$	&$3424.37\pm0.68$&3793
			
			&&\\
			
			&&&&&\\
			
			$1 ^{1}G_{4}$ &$4^{-}$  &$3206.40\pm0.54$	&$3156.85\pm0.39$&
			
			&&&&3364\\
			$2 ^{1}G_{4}$ &$4^{-}$  &$3559.70\pm0.61$	&$3391.65\pm0.54$&

			&&&&&\\
			
			&&&&&\\
			
			$1 ^{3}G_{3}$ &$3^{-}$  &$3213.52\pm0.54$ 	&$3163.26\pm0.39$&
			&&&&3397\\
			$2 ^{3}G_{3}$ &$3^{-}$  &$3565.55\pm0.62$	&$3396.92\pm0.54$&
			&&&&3721\\
			
			&&&&&\\
			
			$1 ^{3}G_{4}$ &$4^{-}$  &$3183.28\pm0.53$ 	&$3135.91\pm0.38$&
			&&&&3399\\
			$2 ^{3}G_{4}$ &$4^{-}$  &$3540.42\pm0.60$	&$3374.34\pm0.53$&
			
			&&&&3722\\
			
			&&&&&\\
			
			$1 ^{3}G_{5}$ &$5^{-}$  &$3139.94\pm0.51$	&$3096.31\pm0.37$&
			&&&&3362\\
			$2 ^{3}G_{5}$ &$5^{-}$  &$3503.53\pm0.58$	&$3340.77\pm0.50$&
			&&&&3685\\
			\hline
		\end{longtable}
	\end{center}			
	
	\subsection{Diquarks}
	Diquarks (\( \mathcal{D} \)) and anti-diquarks (\( \bar{\mathcal{D}} \)) are proposed bound states of two quarks or antiquarks held together by gluonic exchange, and they play a key role in baryon structure according to the quark-diquark model. Although not observed as free particles, diquarks may also serve as the building blocks for exotic hadrons such as tetraquarks and pentaquarks, offering valuable insights into the strong force. When diquarks interact with anti-diquarks, they form composite systems with antisymmetric ground-state wavefunctions due to the Pauli exclusion principle. Depending on their spin structure, diquarks can be scalar (often termed "good diquarks") or axial-vector ("bad diquarks") \cite{Esposito:2016noz}. In the framework of QCD color symmetry, diquarks transform as \(\textbf{3}\otimes\textbf{3} = \bar{\textbf{3}} \oplus \textbf{6}\) \cite{Debastiani:2017msn}, while anti-diquarks follow \(\bar{\textbf{3}}\otimes\bar{\textbf{3}} = \textbf{3} \oplus \bar{\textbf{6}}\). Modeling tetraquarks as bound diquark–anti-diquark systems effectively reduces the four-body problem to a two-body one \cite{Fredriksson:1981mh}. The color coupling \(\bar{\textbf{3}}\otimes\textbf{3}\) results in singlet and octet states, with a color factor \(k_s = -\frac{2}{3}\) that renders the short-range \(\frac{1}{r}\) interaction attractive \cite{Debastiani:2016msc}.
	
	The methodology for calculating diquark and anti-diquark masses parallels that used for mesons. The masses derived in this work incorporate these considerations and are expressed as follows:
	\begin{subequations}
		\begin{equation}
			M_{(qq)} = 2M_{q} + E_{(qq)} + \braket{V^{1}(r)}
		\end{equation} 	
		\begin{equation}
			M_{(Cq)} = M_{C} + M_{q} + E_{(Cq)} + \braket{V^{1}(r)}\\
		\end{equation} 	
	\end{subequations}
	
	where $M_{q}$ and$M_{C}$ are the mass of constituent quarks in the diquark. Table \ref{diquark} shows the mass spectra of calculated diquarks along with comparison with various studies.

	\begin{table*}	
		\centering
		\caption{Mass spectra and comparison  for various Cq diquarks. A comparison with various theoretical models is made. All units are in MeV.}
		\label{diquark}
		\begin{tabular}{ccccccccccc}
			
			\hline
			Mass$_{NR}$ &Mass$_{SR}$ & \cite{Faustov:2021hjs}  &\cite{Chen:2023cws} &\cite{Yin:2021uom} &\cite{Yu:2006ty} &\cite{Giannuzzi:2019esi} &\cite{Gutierrez-Guerrero:2021fuj} &\cite{Tiwari:2022kdu}&\cite{Kleiv:2013dta} &\cite{deOliveira:2023hma} \\ 
			\hline
			$2013.27\pm0.24$  &$2042.04\pm0.25$   &2036 &1920 &2150&2080&2118&1880&1963&1870 &1870 	\\
			\hline
			
		\end{tabular}	
	\end{table*}

	\subsection{Tetraquark Spectra}
	In a color singlet configuration, tetraquark states can be constructed from diquark–antidiquark pairings wherein the diquark resides in a color anti-triplet (\(\bar{\mathbf{3}}\)) and the corresponding antidiquark in a color triplet (\(\mathbf{3}\)) representation. This color configuration, \(\bar{\mathbf{3}}-\mathbf{3}\), is particularly favored due to its attractive nature under one-gluon exchange, characterized by a color factor \(k_s = -\frac{4}{3}\), as detailed in~\cite{Debastiani:2017msn,Debastiani:2016msc}. When both the diquark and antidiquark are in spin-1 states, their coupling can yield a total color singlet tetraquark state, expressible as:
	\[
	\left| QQ\,\middle|^{3} \otimes \bar{Q}\bar{Q} \middle|^{\bar{3}} \right\rangle = \mathbf{1} \oplus \mathbf{8},
	\]
	where only the singlet component (\(\mathbf{1}\)) contributes to a physical color-neutral hadronic state.
	
	Tetraquark systems admit a rich variety of internal configurations, contingent on the specific flavor content and spin alignment of the constituent diquarks and antidiquarks. These configurations significantly affect the total spin \(J\), parity \(P\), and other quantum numbers, as well as the binding energy and stability of the state. Particularly in the light quark sector, the isospin symmetry—while fundamentally approximate—plays an essential role in characterizing these states, despite the minute up-down quark mass difference (\(< 5\,\mathrm{MeV}\)). Hence, isospin is treated as a good approximate quantum number.
	
	In this framework, the mass spectra of singly charmed tetraquarks—comprising one charm quark and three light quarks—are computed for various \(J^P\) configurations using effective quark models.
	\begin{equation}
		M_{Cqqq} =  M_{Cq} + M_{\bar{q}\bar{q}} + E_{(Cq\bar{q}\bar{q})} + \braket{V^{1}(r)}\\		
	\end{equation}
	
	The calculated mass spectra of tetraquark states are influenced by a combination of factors, including the underlying Cornell-like potential, relativistic kinetic energy corrections, and spin-dependent interactions. To isolate and understand the contributions from each component, spin-dependent terms—such as spin-spin, spin-orbit, and tensor interactions—are evaluated separately within a perturbative framework. This methodological separation facilitates a comprehensive analysis of how different physical mechanisms affect the resulting mass hierarchy and spectral structure of tetraquark configurations. Such a systematic approach enhances our understanding of the complex dynamics that govern the behavior of exotic multiquark states.
	
	Furthermore, the color singlet nature of a tetraquark system is achieved through the coupling of the total spin (\(S_T\)) and the total orbital angular momentum (\(L_T\)) of the constituent diquark and antidiquark clusters. This coupling, expressed as \( S_T \otimes L_T \), determines the total angular momentum quantum number \(J_T\) and subsequently the \(J^P\) quantum numbers of the state. These quantum numbers provide crucial insight into the internal structure and spectroscopic properties of tetraquarks, enabling comparisons with experimentally observed candidates and guiding future searches for such states.
	
	\begin{equation}
		\ket{T} = \ket{S_{d},S_{\bar{d}},S_{T},L_{T}}_{J_{T}},
	\end{equation}
	
	Here, the spins of the diquark and anti-diquark are denoted by \( S_{d} \) and \( S_{\bar{d}} \), respectively. In the case of mesons and diquarks, only two spin configurations—namely spin-0 and spin-1—are possible due to the combination of two spin-\(\frac{1}{2}\) constituents. However, for tetraquark systems composed of spin-1 diquarks and spin-1 anti-diquarks, three distinct total spin states can arise from their coupling: \( S_T = 0, 1, \) and \( 2 \). This expanded set of spin configurations highlights the increased structural complexity inherent to tetraquarks, emphasizing the richer internal spin dynamics and the broader spectrum of possible multiquark arrangements allowed by quantum chromodynamics (QCD). The existence of these additional spin states introduces further diversity in the mass spectra and decay patterns of tetraquark candidates, offering valuable avenues for both theoretical exploration and experimental identification.
	
	\begin{equation}
		\begin{split}
			\ket{0^{++}}_{T}=\ket{S_{d}=1,S_{\bar{d}}=1,S_{T}=0,L_{T}=0}_{J_{T=0}};\\
			\ket{1^{+-}}_{T}=\ket{S_{d}=1,S_{\bar{d}}=1,S_{T}=1,L_{T}=0}_{J_{T=1}};\\
			\ket{2^{++}}_{T}=\ket{S_{d}=1,S_{\bar{d}}=1,S_{T}=2,L_{T}=0}_{J_{T=2}};
		\end{split}
	\end{equation}

	The present study adopts a diquark–antidiquark approximation, effectively reducing the four-body problem to a simplified two-body interaction. While this framework offers computational tractability, it inherently limits the ability to incorporate possible mixed-state tetraquark configurations, which may involve more complex correlations beyond the diquark–antidiquark picture. A more comprehensive discussion of this limitation and its implications can be found in~\cite{Debastiani:2017msn}. The computed mass spectra for the singly charmed \( Cq\bar{q}\bar{q} \) tetraquark states in \( S \)-, \( P \)-, and \( D \)-wave configurations are summarized in Table~\ref{mass_tetra}. Furthermore, the corresponding two-meson thresholds relevant for various decay channels are listed in Table~\ref{twomesonthreshold}. The sensitivity of the model to parameter variations—particularly those affecting potential parameters, constituent masses, and spin-dependent couplings—has been systematically investigated in our earlier studies, providing essential insight into the theoretical uncertainties of the predicted spectra \cite{Lodha:2025ffp,Lodha:2024}.
	
	\begin{center}
		
		\begin{longtable}{cccc}
			\label{mass_tetra}\\
			\caption{Mass spectra for singly charm tetraquark for S, P, and D waves in MeV}  \\
			
			\hline
			State &$J^{PC}$ &Mass$_{NR}$ &Mass$_{SR}$ \\
			\toprule
			\endfirsthead
			
			\hline
			State &$J^{PC}$ &Mass$_{NR}$ &Mass$_{SR}$ \\
			\hline
			\endhead
			
			\hline
			\multicolumn{4}{r}{Continued on next page} \\
			\endfoot
			
			\endlastfoot
			
			$1 ^{1}S_{0}$ &	$0^{++}$&$2580.65\pm0.64$ &$2674.99\pm0.63$ \\
			
			$2 ^{1}S_{0}$&$0^{++}$ 	&$3034.34\pm0.54$&$3114.17\pm0.52$ \\
			
			$3 ^{1}S_{0}$&$0^{++}$ 	&$3231.53\pm0.57$&$3308.14\pm0.54$ \\
			
			&&&\\
			$1 ^{3}S_{1}$ &$1^{+-}$&$2652.11\pm0.59$&$2743.73\pm0.58$\\ 
			
			$2 ^{3}S_{1}$&$1^{+-}$&$3053.43\pm0.54$&$3133.04\pm0.52$ \\ 
			
			$3 ^{3}S_{1}$&$1^{+-}$&$3243.79\pm0.57$&$3320.51\pm0.54$ \\

			&&&\\
			$1 ^{5}S_{2}$ &	{$2^{++}$}&$2795.05\pm0.51$&$2876.96\pm0.50$	 \\
			
			$2 ^{5}S_{2}$&{$2^{++}$} &$3091.61\pm0.53$&$3170.43\pm0.52$		\\
			
			$3 ^{5}S_{2}$&{$2^{++}$} &$3268.29\pm0.57$&$3345.09\pm0.54$ \\

			&&&\\
			$1 ^{1}P_{1}$&	$1^{--}$&$3021.69\pm0.53$&$3101.21\pm0.51$ \\
			
			$2 ^{1}P_{1}$&$1^{--}$ &$3209.80\pm0.55$&$3286.74\pm0.53$ \\
			
			$3 ^{1}P_{1}$&$1^{--}$&$3355.91\pm0.59$&$3431.14\pm0.56$ \\
			
			&&&\\
			$1 ^{3}P_{0}$ &	$0^{-+}$&$2956.35\pm0.54$&$3045.26\pm0.52$ \\
			
			$2 ^{3}P_{0}$&$0^{-+}$ &$3159.67\pm0.55$&$3242.77\pm0.53$ \\
			
			$3 ^{3}P_{0}$&$0^{-+}$&$3311.61\pm0.58$&$3391.88\pm0.56$ \\
			
			&&&\\
			
			$1 ^{3}P_{1}$ &	$1^{-+}$&$3020.20\pm0.53$&$3100.47\pm0.51$ \\
			
			$2 ^{3}P_{1}$&$1^{-+}$&$3209.12\pm0.56$&$3286.68\pm0.53$ \\
			
			$3 ^{3}P_{1}$&$1^{-+}$&$3355.43\pm0.59$&$3431.19\pm0.56$ \\
			
			&&&\\
			
			$1 ^{3}P_{2}$&$2^{-+}$&$3042.43\pm0.53$&$3119.24\pm0.51$ \\
			
			$2 ^{3}P_{2}$&$2^{-+}$&$3226.58\pm0.56$ &$3301.82\pm0.54$ \\

			$3 ^{3}P_{2}$&$2^{-+}$&$3371.07\pm0.6$ 	&$3444.99\pm0.57$  \\

			&&&\\
			$1 ^{5}P_{1}$ &	$1^{--}$&$2953.36\pm0.54$ &$3043.77\pm0.52$  \\
			
			$2 ^{5}P_{1}$&$1^{--}$&$3158.32\pm0.55$ &$3242.56\pm0.53$  \\
			
			$3 ^{5}P_{1}$&$1^{--}$&$3310.64\pm0.58$ &$3391.99\pm0.56$ \\
			
			&&&\\
			$1 ^{5}P_{2}$ &$2^{--}$&$2952.73\pm0.54$ &$3042.93\pm0.52$ \\
			
			$2 ^{5}P_{2}$&$2^{--}$&$3157.97\pm0.55$	&$3242.06\pm0.53$ \\
			
			$3 ^{5}P_{2}$&$2^{--}$&$3310.44\pm0.58$	&$3391.66\pm0.56$ \\
			
			&&&\\
			$1 ^{5}P_{3}$&	$3^{--}$&$3033.59\pm0.53$&$3112.65\pm0.51$ 			\\
			
			$2 ^{5}P_{3}$&$3^{--}$&$3220.7\pm0.56$	&$3297.58\pm0.54$ 		\\
			
			$3 ^{5}P_{3}$&$3^{--}$&$3366.11\pm0.6$	&$3441.49\pm0.57$ 			\\
			
			&&&\\
			
			$1 ^{1}D_{2}$ &$2^{++}$&$3165.66\pm0.54$ &$3242.44\pm0.53$ \\
			
			$2 ^{1}D_{2}$&	$2^{++}$&$3315.92\pm0.58$ 	&$3391.44\pm0.55$ \\
			
			$3 ^{1}D_{2}$&	$2^{++}$&$3444.36\pm0.62$ 	&$3518.48\pm0.59$ 	\\
			
			&&&\\
		
			$1 ^{3}D_{1}$&	$1^{+-}$&$3154.7\pm0.54$ 	&$3232.98\pm0.52$\\
		
			$2 ^{3}D_{1}$&	$1^{+-}$&$3305.71\pm0.58$ 	&$3382.49\pm0.55$ \\
		
			$3 ^{3}D_{1}$&	$1^{+-}$&$3434.55\pm0.62$ &$3509.83\pm0.58$			\\
			
			&&&\\
			
			$1 ^{3}D_{2}$&	$2^{+-}$&$3164.09\pm0.54$ &$3241.19\pm0.53$ \\
			
			$2 ^{3}D_{2}$&	$2^{+-}$&$3314.5\pm0.58$ &$3390.29\pm0.55$ 		\\
			
			$3 ^{3}D_{2}$&	$2^{+-}$&$3443.03\pm0.62$ 	&$3517.43\pm0.59$			\\
			
			&&&\\
			
			$1 ^{3}D_{3}$&	$3^{+-}$&$3171.00\pm0.54$ &$3247.00\pm0.53$\\
			
			$2 ^{3}D_{3}$&	$3^{+-}$&$3321.18\pm0.58$ & $3396.04\pm0.55$\\
			
			$3 ^{3}D_{3}$&	$3^{+-}$&$3449.61\pm0.62$ &$3523.17\pm0.59$			\\
			
			&&&\\
			
			$1 ^{5}D_{0}$&	$0^{++}$&$3148.24\pm0.54$ &$3227.62\pm0.53$			\\
			
			$2 ^{5}D_{0}$&	$0^{++}$&$3299.72\pm0.58$ &$3377.44\pm0.55$			\\
			
			$3 ^{5}D_{0}$&	$0^{++}$&$3428.82\pm0.61$ &$3504.97\pm0.58$			\\
			
			&&&\\
			
			$1 ^{5}D_{1}$&	$1^{++}$&$3144.17\pm0.54$ &$3223.94\pm0.52$		\\
			
			$2 ^{5}D_{1}$&	$1^{++}$&$3296.04\pm0.57$ &$3374.07\pm0.55$		\\
			
			$3 ^{5}D_{1}$&	$1^{++}$&$3425.34\pm0.61$ &$3501.77\pm0.58$\\
			
			&&&\\
			
			$1 ^{5}D_{2}$&	$2^{++}$&$3149.82\pm0.54$ &$3228.91\pm0.53$\\
			
			$2 ^{5}D_{2}$&	$2^{++}$&$3301.3\pm0.58$ &$3378.76\pm0.55$\\
			
			$3 ^{5}D_{2}$&	$2^{++}$&$3430.4\pm0.61$ &$3506.62\pm0.58$	\\
			
			&&&\\
			
			$1 ^{5}D_{3}$&	$3^{++}$&$3159.18\pm0.54$ &$3237.08\pm0.51$\\
			
			$2 ^{5}D_{3}$&	$3^{++}$&$3310.05\pm0.58$ &$3386.53\pm0.55$		\\
			
			$3 ^{5}D_{3}$&	$3^{++}$&$3438.85\pm0.62$ &$3513.89\pm0.59$		\\
			
			&&&\\
			
			$1 ^{5}D_{4}$&	$4^{++}$&$3169.53\pm0.54$ &$3247.51\pm0.52$ \\
			
			$2 ^{5}D_{4}$&	$4^{++}$&$3319.84\pm0.58$ &$3395.44\pm0.57$ 		\\
			
			$3 ^{5}D_{4}$&	$4^{++}$&$3448.37\pm0.62$ &$3521.51\pm0.65$		\\
			
			\hline
		\end{longtable}
	\end{center}

	\begin{longtable}{cccc}
		\caption{Two-meson threshold for different tetraquark states in MeV} \label{twomesonthreshold} \\
		
		\toprule
		\multirow{3}{*}{State} & \multirow{2}{30mm}{Two-meson Threshold} 
		& \multicolumn{2}{c}{Threshold Mass}\\
		& & Non-Relativistic & Semi-Relativistic \\   \\
		
		\midrule
		\endfirsthead
		
		\toprule
		\multirow{3}{*}{State} & \multirow{2}{30mm}{Two-meson Threshold} 
		& \multicolumn{2}{c}{Threshold Mass}\\
		& & Non-Relativistic & Semi-Relativistic \\   \\
		\midrule
		\endhead
		
		\multicolumn{3}{r}{Continued on next page...} \\
		\endfoot
		
		\bottomrule
		\endlastfoot
		
		$^{1}S_{0}$ &$\pi(1S)\; D(1S)$ &2006.98 &2002.14 \\
		\midrule
		
		\multirow{2}{*}{$^{3}S_{1}$} &$\pi(1S)\; D^{*}(1S)$ &2146.94 &2146.26 \\
		&$\rho(1S)\; D(1S)$ &2640.19 &2637.41 \\
		\midrule
		
		$^{5}S_{2}$  &$\rho(1S) \; D^{*}(1S)$ &2780.15 &2781.53 \\
		\midrule
		
		\multirow{2}{*}{$^{1}P_{1}$} &$\pi(1S)D(1P)$&2551.84 & 2522.70 \\
		
		&$D(1S)b_{1}(1P)$&3129.05 & 3141.19 \\
		
		\midrule
		
		\multirow{2}{*}{$^{3}P_{0}$} &$\pi(1S)$  $D^{*}_{0}(1P)$ &2479.78 & 2493.57 \\
		
		& $D(1S)$  $a_{0}(1P)$ &2728.19 & 2765.97 \\
		
		\midrule
		
		\multirow{2}{*}{$^{3}P_{1}$} & $\pi(1S)$  $D^{*}_{1}(1P)$ &2581.60 & 2595.82 \\
		& $D(1S)$  $a_{1}(1P)$ &3157.07 & 3178.65 \\
		\hline
		
		\multirow{2}{*}{$^{3}P_{2}$} & $\pi(1S)$  $D^{*}_{2}(1P)$ &	2631.51 & 2648.02 \\
		& $D(1S)$  $a_{2}(1P)$ &				3200.96 & 3232.61 \\
		\hline
		
		\multirow{2}{*}{$^{5}P_{1}$} & $\rho(1S)$  $D^{*}_{0}(1P)$ &				3112.99 & 3128.84 \\
		& $D^{*}(1S)$  $a_{0}(1P)$&				3874.72 & 3869.42 \\
		\hline
		
		\multirow{2}{*}{$^{5}P_{2}$} & $\rho(1S)$  $D^{*}_{1}(1P)$ &				3214.81 & 3231.09 \\
		& $D^{*}(1S)$  $a_{1}(1P)$ &				3297.03 & 3322.77 \\
		\hline
		
		\multirow{2}{*}{$^{5}P_{3}$} & $\rho(1S)$  $D^{*}_{2}(1P)$ &				3264.72 & 3283.29 \\
		& $D^{*}(1S)$  $a_{2}(1P)$ &				3340.92 & 3376.73 \\
		\midrule
		
		\multirow{2}{*}{$^{1}D_{2}$} &$\pi(1S) \; D_{2} (1D)$ &				2861.45 & 2866.14 \\
		&$D(1S) \; \pi_{2}(1D)$ &				3491.91 & 3488.58 \\
		\midrule
		
		\multirow{2}{*}{$^{3}D_{1}$} &$\pi(1S) \; D^{*}_{1} (1D)$ &				2857.41 & 2862.12 \\
		&$D(1S) \; \rho_{1}(1D)$ &				3438.77 & 3444.13 \\
		\midrule
		
		\multirow{2}{*}{$^{3}D_{2}$} &$\pi(1S) \; D^{*}_{2} (1D)$ &				2866.62 & 2872.93 \\
		&$D(1S) \; \rho_{2}(1D)$ &				3473.81 & 3489.19 \\
		\midrule
		
		\multirow{2}{*}{$^{3}D_{3}$} &$\pi(1S) \; D^{*}_{3} (1D)$ &				2855.20 & 2864.34 \\
		&$D(1S) \; \rho_{3}(1D)$ &				3420.23 & 3506.65 \\
		\midrule
		
		\multirow{2}{*}{$^{5}D_{0}$} &$\rho(1S) \; D^{*}_{1} (1D)$ &				3490.62 & 3497.39 \\
		&$D^{*}(1S) \; \rho_{1}(1D)$ &				3578.73 & 3588.25 \\
		\midrule
		
		\multirow{2}{*}{$^{5}D_{1}$} &$a_{0}(1P) \; D^{*}_{1} (1D)$ &				3578.62 & 3625.95 \\
		&$D^{*}(1P) \; \rho_{1}(1D)$ &				3911.57 & 3935.56 \\
		\midrule
		
		\multirow{2}{*}{$^{5}D_{2}$} &$\rho(1S) \; D^{*}_{1} (1D)$ &				3490.62 & 3497.39 \\
		&$D^{*}(1S) \; \rho_{1}(1D)$ &				3578.73 & 3588.25 \\
		\midrule
		\multirow{2}{*}{$^{5}D_{3}$}&$\rho(1S) \; D^{*}_{2} (1D)$ &				3499.83 & 3508.20 \\
		&$D^{*}(1S) \; \rho_{2}(1D)$ &				3613.77 & 3633.31 \\
		\midrule
		\multirow{2}{*}{$^{5}D_{4}$}&$\rho(1S) \; D^{*}_{3} (1D)$ &				3488.41 & 3499.61 \\
		&$D^{*}(1S) \; \rho_{3}(1D)$ &				3560.19 & 3650.77 \\
	\end{longtable}

	\section{Decay}
	\label{sec:3}
	A fundamental property accompanying the mass spectra of bound states is their decay width, which offers critical insight into the stability and underlying dynamics of hadronic systems. In this context, \( D \) mesons—decaying predominantly via weak interactions—serve as sensitive probes for exploring the structure of weak processes. Nonleptonic decays of heavy mesons, in particular, provide a valuable platform to investigate the interplay between weak and strong interactions. While numerous such decay channels have been measured experimentally, their theoretical understanding remains incomplete and necessitates the development of robust, systematically reliable approaches.
	
	To analyze the nonleptonic decays of heavy mesons, the present study adopts the factorization hypothesis, wherein the decay amplitude is decomposed into a product of a meson decay constant and weak transition form factors \cite{Bjorken:1988kk,Shifman:1991vm}. This method facilitates the separation of short-distance contributions—amenable to treatment via perturbative QCD—from long-distance hadronic effects, which are encapsulated in the form factors and require nonperturbative techniques or phenomenological modeling.
	
	In parallel, the decay mechanisms of tetraquarks—exotic four-quark bound states—pose challenges comparable to those encountered in their mass spectrum calculations. In the present framework, the hadronic decay processes of tetraquarks are addressed using a quark rearrangement model, which accounts for the recombination of constituent quarks into two-meson final states \cite{Lodha:2024bwn,Lodha:2024qby}. This approach captures essential features of strong decays and provides a tractable method for estimating decay patterns and widths within the tetraquark sector.
	
	\subsection{Meson Decay}
	
	Decays of \( D \) mesons provide critical insight into the interplay between weak and strong interactions in systems composed of both heavy and light quarks. Their analysis is essential not only for testing the Standard Model but also for probing potential avenues of new physics. Among the various decay processes, nonleptonic channels are especially complex due to the involvement of strongly interacting final-state hadrons, leading to nontrivial hadronic matrix elements.
	
	To address this complexity, the factorization approach is employed, wherein the hadronic matrix elements are simplified by decomposing them into products involving meson decay constants and weak transition form factors. This method separates the perturbatively calculable short-distance contributions from the nonperturbative long-distance dynamics. A key component of this analysis is the use of form factors derived from a relativistic quark model based on the quasi-potential approach. These form factors have been validated against experimental data from semileptonic decays of \( D \) mesons, showing strong consistency. By utilizing these well-established form factors, the study ensures an accurate treatment of long-distance hadronic effects in nonleptonic \( D \) meson decays \cite{Faustov:2019mqr}.
	
	The effective Hamiltonian governing the weak decays of \( D \) mesons is given by:
	\begin{equation}
		{H}_{\text{eff}} = \frac{G_{F}}{\sqrt{2}} V_{cq_1}^{*} V_{uq_2} \left[ 
		c_1(\mu)(\bar{q}_{1\alpha} c_{\alpha})_{V-A}(\bar{u}_{\beta} q_{2\beta})_{V-A} + 
		c_2(\mu)(\bar{q}_{1\alpha} c_{\beta})_{V-A}(\bar{u}_{\beta} q_{2\alpha})_{V-A} \right],
	\end{equation}
	where \( q_1 = q_2 = d \), \( G_F \) is the Fermi coupling constant, \( \alpha \) and \( \beta \) are color indices, and \( c_1 \), \( c_2 \) are the Wilson coefficients, taken as \( c_1 = 1.26 \) and \( c_2 = -0.51 \) at the scale \( \mu = m_c \) \cite{Kamal:1995fr}.
	
	Using Fierz transformations and Gell-Mann matrix identities, the effective Hamiltonian can be re-expressed in terms of color-favored and color-suppressed parts:
	\begin{subequations}
		\begin{align}
			{H}_{\text{cf}} &= \frac{G_{F}}{\sqrt{2}} V_{cq_1}^{*} V_{uq_2} \, a_1 (\bar{q}_1 c)_{V-A} (\bar{u} q_2)_{V-A}, \\
			{H}_{\text{cs}} &= \frac{G_{F}}{\sqrt{2}} V_{cq_1}^{*} V_{uq_2} \, a_2 (\bar{q}_1 q_2)_{V-A} (\bar{u} c)_{V-A},
		\end{align}
	\end{subequations}
	where the effective coefficients are defined as $a_1 = c_1 + \frac{c_2}{N}, \quad a_2 = c_2 + \frac{c_1}{N}$. Here, \( N \) is the number of colors, and in the large \( N \) limit, \( N \to \infty \), one obtains \( a_1 \approx c_1 \) and \( a_2 \approx c_2 \). Within the factorization framework, the hadronic matrix elements can then be expressed as products of decay constants and invariant form factors, allowing for a tractable and systematic treatment of nonleptonic decay processes. The decay constant is defined as the matrix element of the weak current between the vacuum and a pseudoscalar (P) or a vector (V) meson:
	
	\begin{subequations}
		\begin{equation}
			\bra{P(p_{\mu})}(\bar{q_{1}}q_{2})_{V-A}\Ket{0}=-if_{P}p_{\mu}
		\end{equation} 
		\begin{equation}
			\bra{V}(\bar{q_{1}}q_{2})_{V-A}\Ket{0}=if_{V}m_{V} \epsilon^{*}_{\mu}
		\end{equation}
	\end{subequations}
	where $f_{P}$ and $f_{V}$ are the decay constants of the pseudoscalar and vector meson, respectively; $m_{V}$ and $\epsilon_{mu}$ are the mass and polarization vector of the vector meson. In the rest frame of the initial meson, one has the $D$ explicit representations of the momentum and polarization vectors:
	\begin{equation}
		\begin{split}
			p^{\mu}_{D} = (m_{D},0,0,0)\\
			p^{\mu}_{M_{1}} = (E_{1},0,0,|\textbf{p}|)\\
			q^{\mu}= (E_{2},0,0,\textbf{-}|\textbf{p}|)\\
			\epsilon^{\dagger\mu}_{t} = \frac{1}{\sqrt{q^{2}}}(E_{2},0,0,\textbf{-}|\textbf{p}|)\\
			\epsilon^{\dagger\mu}_{\pm} = \frac{1}{\sqrt{2}}(0,\pm1,i,0)\\
			\epsilon^{\dagger\mu}_{0} = \frac{1}{\sqrt{q^{2}}}(|\textbf{p}|,0,0,\textbf{-}E_{2})\\
		\end{split}
	\end{equation} 
	where $E_{1}$ is the energy of $M_{1}$ and $E_{1}+E_{2}=m_{D}$ , $|\textbf{p}|=\lambda^{1/2}(m^{2}_{D},m^{2}_{1},m^{2}_{2})/(2m_{D})$ is the momentum of the
	daughter meson with $\lambda(x,y,z)=x^{2}+y^{2}+z^{2}-2(xy+yz+xz)$. The transition matrix for various channels is given as \cite{Zhang:2020dla},
	
	\begin{equation}
		\begin{split}
			\mathcal{M}_{D\rightarrow P_{1}P_{2}} &=-if_{P_{2}}m_{P_{2}}\epsilon_{t}^{\dagger\mu}\bra{P_{1}}(\bar{q}c)_{V-A}\ket{D}\\
			&=-if_{P_{2}}m_{P_{2}}H_{t}
		\end{split}
	\end{equation} 
	\begin{equation}
		\begin{split}
			\mathcal{M}_{D\rightarrow P,V} &=-if_{V}m_{V}\epsilon_{\lambda}^{\dagger\mu}\bra{P}(\bar{q}c)_{V-A}\ket{D}\\
			&=if_{V}m_{V}H_{0}
		\end{split}
	\end{equation}
	\begin{equation}
		\begin{split}
			\mathcal{M}_{D\rightarrow V,P} &=-if_{P}m_{P}\epsilon_{t}^{\dagger\mu}\bra{V}(\bar{q}c)_{V-A}\ket{D}\\
			&=-if_{P}m_{P}H_{t}
		\end{split}
	\end{equation}
	\begin{equation}
		\begin{split}
			|\mathcal{M}_{D\rightarrow V_{1},V_{2}}|^{2} &=f_{V}^{2}m_{V}^{2} (|H_{+}|^{2}+|H_{-}|^{2}+|H_{0}|^{2})\\
		\end{split}
	\end{equation}
	where,
	\begin{equation}
		H_{\pm}=-(m_{D}+m_{V1})A_{1}(m_{V_{2}}^{2})\pm \frac{2m_{D}|\textbf{p}|}{m_{D}+m_{V_{1}}}V(m^{2}_{V_{2}})
	\end{equation}
	\begin{equation}
		H_{0}=-(m_{D}+m_{V1})A_{1}(m_{V_{2}}^{2})\frac{m^{2}_{D}-m_{V_{1}}^{2}-m^{2}_{V_{2}}}{2m_{V_{1}}m_{V_{2}}} + \frac{2m_{D}^{2}|\textbf{p}|^{2}}{(m_{D}+m_{V_{1}})m_{V_{1}}m_{V_{2}}}A_{2}(m^{2}_{V_{2}})
	\end{equation}
	
	\begin{center}
		\begin{table*}
			
			\centering
			\caption{Decay constants of various mesons in MeV}
			\label{decay_constant}
			\begin{tabular}{cccccccccc}
				\hline
				$f_{\pi}$ & $f_{K}$ &$f_{K^{*}}$  &$f_{\eta}^{u}$  &$f_{\eta}^{s}$  &$f_{\eta^{'}}^{u}$  &$f_{\eta^{'}}^{s}$  &$f_{\rho}$  &$f_{\omega}$  &$f_{\phi}$   \\
				\hline
				130 & 156 & 217 & 78 & -112 & 63 & 137 & 205 & 187 & 215    \\
				\hline
			\end{tabular}
		\end{table*}
	\end{center}
	
	The values of the decay constants used in our calculations are tabulated in table \ref{decay_constant}. The relevant form factor are expressed as function of momentum tranfer squared. In case of $f_{+}(q^{2}),V(q^{2})$ and $A_{0}(q^{2})$, the expression follows :
	\begin{equation}
		F(q^{2})=\frac{F(0)}{\biggr(1-\frac{q^{2}}{M^{2}}\biggr)\biggr[1-\sigma_{1} \frac{q^{2}}{M_{1}^{2}}+\sigma_{2}\biggr(\frac{q^{2}}{M_{1}^{2}}\biggr)^{2}\biggr]}
	\end{equation}
	Similarly, $f_{0}(q^{2}),A_{1}(q^{2})$ and $A_{2}(q^{2})$ are expressed as,
	\begin{equation}
		F(q^{2})=\frac{F(0)}{1-\sigma_{1} \frac{q^{2}}{M_{1}^{2}}+\sigma_{2}\biggr(\frac{q^{2}}{M_{1}^{2}}\biggr)^{2}}
	\end{equation}
	
	For the \( c \rightarrow d \) transition, the mass \( M \) used in the form factors \( f_{+}(q^{2}) \) and \( V(q^{2}) \) corresponds to the mass of the pseudoscalar \( D \) meson in its ground state (\( 1^{1}S_{0} \)). Similarly, for the form factor \( A_{0}(q^{2}) \), \( M \) is also taken as the mass of the \( 1^{1}S_{0} \) \( D \) meson. The parameter \( M_1 \) denotes the pole mass between the active quarks, and is identified with the mass of the vector meson \( D^{*} \), i.e., \( M_{D^{*}} \). A detailed explanation of the factorization approach and the construction of form factors is provided in Ref.~\cite{Yu:2022ngu}.
	
	The differential branching fraction for a nonleptonic decay of the \( D \) meson into final states \( M_1 \) and \( M_2 \) is given by:
	\begin{equation}
		\frac{d\mathcal{B}}{dq^{2}} = \frac{\tau_D}{512 \pi^{3} M_D^{3}} \left| \mathcal{M}(D \rightarrow M_1 M_2) \right|^2 \lambda^{1/2}(M_D^{2}, M_1^{2}, M_2^{2}),
	\end{equation}
	where \( \tau_D \) is the lifetime of the \( D \) meson, \( \mathcal{M} \) is the decay amplitude obtained using the factorization hypothesis, and \( \lambda(a,b,c) = a^2 + b^2 + c^2 - 2ab - 2ac - 2bc \) is the Källén function representing the triangle function from phase space considerations.
	
	For practical calculations, the amplitude \( \mathcal{M} \) is constructed using the effective Hamiltonians \( \mathcal{H}_{\text{cf}} \) and \( \mathcal{H}_{\text{cs}} \), and the corresponding hadronic matrix elements expressed through the decay constants and form factors. The branching fraction is given by,
	\begin{equation}
		\mathcal{B}=\tau\frac{|\textbf{p}|}{8\pi m^{2}}|A|^{2}
	\end{equation}
	where, $\tau$and $m$ are the lifetime and mass of D meson, respectively. |$\textbf{p}$| is the magnitude of three momentum of the final meson in rest frame of heavy meson and the expression of amplitude is given by A. The branching ratios for various decay channel for $D$ meson in present work and other theoretical studies are tabulated in table \ref{Branch}. 
	
	It should be noted that the factorization approach neglects non-factorizable contributions such as soft-gluon exchanges between final-state mesons. These effects may become relevant in certain decay modes and typically require model-dependent treatments or lattice QCD corrections. Moreover, final-state interactions (FSIs) can modify decay amplitudes through rescattering processes, which are not accounted for in the naive factorization model but may play a phenomenologically significant role. Many of the decay constants and form factor parameters used in this approach are constrained using experimental measurements of semileptonic and leptonic decays, thereby ensuring consistency with observed decay rates. While the factorization approach offers a practical and widely adopted method for estimating nonleptonic decay amplitudes, its limitations underscore the importance of incorporating complementary techniques such as QCD sum rules, lattice QCD, or phenomenological models to account for non-factorizable effects. Continued progress in both theoretical modeling and experimental precision will be crucial for deepening our understanding of charm quark dynamics and hadronic decay mechanisms.

	\begin{center}
		\begin{table*}
			
			\caption{Branching fractions for $D$ decays and their comparison with theoretical studies}
			\label{Branch}
		\resizebox{\textwidth}{!}{
			{\begin{tabular}{cccccccc}
					\toprule
					Decay Channel &Non Relativistic &Semi Relativistic &\cite{ParticleDataGroup:2024cfk}     & \cite{Manak} & \cite{Manak}  & \cite{Yu:2022ngu} & \cite{Biswas:2015aaa} \\
					\midrule
					$ \pi^0 \pi^+$ &2.54$\times 10^{-3}$ & 2.52$\times 10^{-3}$ &$1.25 \times 10^{-3}$ & $1.91 \times 10^{-3}$ & $2.53 \times 10^{-3}$ & $1.3 \times 10^{-3}$ & $8.9 \times 10^{-4}$  \\

					$ \pi^0 K^+$ &4.86$\times 10^{-4}$ & 4.83$\times 10^{-4}$ &
					$2.08 \times 10^{-4}$ & $3.6 \times 10^{-4}$ & $4.8 \times 10^{-4}$ & $2.5 \times 10^{-4}$ & $3.1 \times 10^{-4}$ \\
					
					$ \eta K^+$ & 2.90$\times 10^{-4}$ & 2.88$\times 10^{-4}$ & $1.3 \times 10^{-4}$ & $2.0 \times 10^{-4}$ & $2.9 \times 10^{-4}$ & $3.0 \times 10^{-4}$ & $1.0 \times 10^{-4}$ \\

					$ \eta^{\prime} K^+$ & 1.16$\times 10^{-4}$ & 1.15$\times 10^{-4}$ & $1.9 \times 10^{-4}$ & $6.0 \times 10^{-5}$ & $1.2 \times 10^{-4}$ & $1.2 \times 10^{-4}$ & $1.4 \times 10^{-4}$ \\

					$ \eta \pi^+$ & 3.03$\times 10^{-4}$ & 3.00$\times 10^{-4}$ & $3.77 \times 10^{-3}$ & $2.0 \times 10^{-4}$ & $3.0 \times 10^{-4}$ & $3.1 \times 10^{-4}$ & $4.72 \times 10^{-3}$ \\

					$ \eta^{\prime} \pi^+$ & 3.57$\times 10^{-3}$ & 3.53$\times 10^{-3}$ & $4.97 \times 10^{-3}$ & $2.22 \times 10^{-3}$ & $3.55 \times 10^{-3}$ & $3.66 \times 10^{-3}$ & $6.76 \times 10^{-3}$ \\
					
					$ \pi^+ \rho^0$ & 4.75$\times 10^{-4}$ & 4.71$\times 10^{-4}$ & $8.4 \times 10^{-4}$ & $3.1 \times 10^{-4}$ & $4.7 \times 10^{-4}$ & $2.4 \times 10^{-4}$ & - \\

					$ \pi^+ \phi$ & 2.17$\times 10^{-3}$ & 2.15$\times 10^{-3}$ & $5.7 \times 10^{-3}$ & $1.22 \times 10^{-3}$ & $2.16 \times 10^{-3}$ & $2.23 \times 10^{-3}$ & - \\

					$ \pi^+ \omega$ & 3.74$\times 10^{-4}$ & 3.71$\times 10^{-4}$ & $2.8 \times 10^{-4}$ & $2.5 \times 10^{-4}$ & $3.7 \times 10^{-4}$ & $1.9 \times 10^{-4}$ & - \\
					
					$ \rho^0 K^+$ & 3.18$\times 10^{-4}$ & 3.14$\times 10^{-4}$ & $1.9 \times 10^{-4}$ & $1.8 \times 10^{-4}$ & $3.2 \times 10^{-4}$ & $1.7 \times 10^{-4}$ & - \\
					
					$ \rho^+ \phi$ & 2.65$\times 10^{-3}$ & 2.57$\times 10^{-3}$ & $<1.5 \times 10^{-2}$ & $2.48 \times 10^{-3}$ & $2.60 \times 10^{-3}$ & $2.74 \times 10^{-3}$ & - \\

					$ K^- \pi^+$ & 5.48$\times 10^{-2}$ & 5.44$\times 10^{-2}$ & $3.947 \times 10^{-2}$ & $3.96 \times 10^{-2}$ & $5.46 \times 10^{-2}$ & $5.44 \times 10^{-2}$ & $3.7 \times 10^{-2}$ \\

					$ \pi^- \pi^+$ & 2.84$\times 10^{-3}$ & 2.82$\times 10^{-3}$ & $1.45 \times 10^{-3}$ & $2.14 \times 10^{-3}$ & $2.83 \times 10^{-3}$ & $2.86 \times 10^{-3}$ & $1.44 \times 10^{-3}$ \\

					$ \pi^0 \pi^0$ & 1.86$\times 10^{-3}$ & 1.85$\times 10^{-3}$ & 
					$8.3 \times 10^{-4}$ & $1.4 \times 10^{-3}$ & $1.85 \times 10^{-3}$ & $2.4 \times 10^{-4}$ & $1.14 \times 10^{-3}$ \\
					
					$ K^- K^+$ & 3.94$\times 10^{-3}$ & 3.91$\times 10^{-3}$ & $4.08 \times 10^{-3}$ & $2.78 \times 10^{-3}$ & $3.92 \times 10^{-3}$ & $3.96 \times 10^{-3}$ & $4.06 \times 10^{-3}$ \\

					$ \eta \eta$ & 2.06$\times 10^{-3}$ & 2.04$\times 10^{-3}$ & $2.11 \times 10^{-3}$ & $1.42 \times 10^{-3}$ & $2.05 \times 10^{-3}$ & $2.07 \times 10^{-3}$ & $1.27 \times 10^{-3}$ \\

					$ \pi^- K^+$ & 1.93$\times 10^{-4}$ & 1.91$\times 10^{-4}$ & $1.5 \times 10^{-4}$ & $1.4 \times 10^{-4}$ & $1.9 \times 10^{-4}$ & $2.0 \times 10^{-4}$ & $1.8 \times 10^{-4}$ \\

					$ \eta \pi^0$ & 2.07$\times 10^{-3}$ & 2.06$\times 10^{-3}$ & $6.3 \times 10^{-4}$ & $1.51 \times 10^{-3}$ & $2.07 \times 10^{-3}$ & $9.0 \times 10^{-5}$ & $1.47 \times 10^{-3}$ \\

					$ \eta^{\prime} \pi^0$ & 4.96$\times 10^{-6}$ & 4.85$\times 10^{-6}$ & $9.2 \times 10^{-4}$ & $1.4 \times 10^{-6}$ & $4.9 \times 10^{-6}$ & $2.2 \times 10^{-4}$ & $2.17 \times 10^{-3}$ \\

					$ \eta \eta^{\prime}$ & 4.97$\times 10^{-5}$ & 4.89$\times 10^{-5}$ & $1.01 \times 10^{-3}$ & $2.0 \times 10^{-5}$ & $5.0 \times 10^{-5}$ & $5.0 \times 10^{-5}$ & $9.5 \times 10^{-4}$ \\

					$ \pi^0 \omega$ & 2.01$\times 10^{-3}$ & 1.99$\times 10^{-3}$ & $1.2 \times 10^{-4}$ & $1.32 \times 10^{-3}$ & $1.99 \times 10^{-3}$ & $4.0 \times 10^{-5}$ & - \\

					$ \eta \omega$ & 1.79$\times 10^{-3}$ & 1.76$\times 10^{-3}$ & $1.98 \times 10^{-3}$ & $9.5 \times 10^{-4}$ & $1.77 \times 10^{-3}$ & $9.0 \times 10^{-4}$ & - \\

					$ \rho^0 \pi^0$ & 2.43$\times 10^{-3}$ & 2.41$\times 10^{-3}$ & $3.86 \times 10^{-3}$ & $1.6 \times 10^{-3}$ & $2.42 \times 10^{-3}$ & $6.1 \times 10^{-4}$ & - \\

					$ \pi^- \rho^+$ & 5.92$\times 10^{-3}$ & 5.87$\times 10^{-3}$ & $1.01 \times 10^{-2}$ & $3.89 \times 10^{-3}$ & $5.88 \times 10^{-3}$ & $5.95 \times 10^{-3}$ & - \\

					$ \pi^0 \phi$ & 8.65$\times 10^{-4}$ & 8.55$\times 10^{-4}$ & 
					$1.17 \times 10^{-3}$ & $4.9 \times 10^{-4}$ & $8.6 \times 10^{-4}$ & $4.3 \times 10^{-4}$ & - \\
					
					$ \rho^- \pi^+$ & 2.01$\times 10^{-3}$ & 2.00$\times 10^{-3}$ & 
					$5.15 \times 10^{-3}$ & $1.32 \times 10^{-3}$ & $2.0 \times 10^{-3}$ & $2.02 \times 10^{-3}$ & - \\
					
					$ \eta \phi$ & 3.88$\times 10^{-4}$ & 3.80$\times 10^{-4}$ & 
					$1.8 \times 10^{-4}$ & $1.1 \times 10^{-4}$ & $3.8 \times 10^{-4}$ & $3.9 \times 10^{-4}$ & - \\
					
					$ K^- \rho^+$ & 1.07$\times 10^{-1}$ & 1.06$\times 10^{-1}$ & 
					$1.12 \times 10^{-1}$ & $6.05 \times 10^{-2}$ & $1.06 \times 10^{-1}$ & $1.06 \times 10^{-1}$ & - \\
					
					$ \eta \bar{K}^{*0}$ & 9.91$\times 10^{-3}$ & 9.76$\times 10^{-3}$ & 
					$1.41 \times 10^{-2}$ & $4.32 \times 10^{-3}$ & $9.81 \times 10^{-3}$ & $9.86 \times 10^{-3}$ & - \\
					
					$ \eta^{\prime} \bar{K}^{*0}$ & 6.60$\times 10^{-5}$ & 4.04$\times 10^{-5}$ & 
					$<1.0 \times 10^{-3}$ & $2.27 \times 10^{-3}$ & $5.0 \times 10^{-5}$ & $8.0 \times 10^{-5}$ & - \\
					
					$ \rho^0 \rho^0$ & 5.53$\times 10^{-3}$ & 5.47$\times 10^{-3}$ & 
					$1.85 \times 10^{-3}$ & $3.2 \times 10^{-3}$ & $5.49 \times 10^{-3}$ & $6.9 \times 10^{-4}$ & - \\
					
					$ \omega \phi$ & 9.54$\times 10^{-4}$ & 9.22$\times 10^{-4}$ & 
					$6.5 \times 10^{-4}$ & $9.7 \times 10^{-4}$ & $9.3 \times 10^{-4}$ & $4.7 \times 10^{-4}$ & - \\
					\bottomrule& 
			\end{tabular}}
}
		\end{table*}
	\end{center}

	\subsection{Tetraquark Deacy}
	
	The recoupling of spin wave functions is given by:
	\begin{subequations}
		\begin{equation}		
			\bigr|{\{{(Cq)}^{1}{(\bar{q}\bar{q})}^{1}\}^{0}}\bigr \rangle=-\frac{1}{2} \bigr|\{\bigr(C\bar{q}\bigr)^{1}\bigr(q\bar{q}\bigr)^{1}\bigr \}^{0}\rangle  + \frac{\sqrt{3}}{2}\bigr|\{\bigr(C\bar{q}\bigr)^{0}\bigr(q\bar{q}\bigr)^{0}\bigr\}^{0} \rangle  	
			\label{spinwfsqqqJ0}		
		\end{equation}
		\begin{equation}		
			\bigr|{\{{(Cq)}^{1}{(\bar{q}\bar{q})}^{1}\}^{1}}\bigr \rangle=\frac{1}{\sqrt{2}} \bigr|\{\bigr(C\bar{q}\bigr)^{1}\bigr(q\bar{q}\bigr)^{0}\bigr\}^{1} \rangle  + \frac{{1}}{\sqrt{2}}\bigr|\{\bigr(C\bar{q}\bigr)^{0}\bigr(q\bar{q}\bigr)^{1}\}^{1}\bigr \rangle  	
			\label{spinwfsqqqJ1}		
		\end{equation}
		\begin{equation}		
			\bigr|{\{{(Cq)}^{1}{(\bar{q}\bar{q})}^{1}\}^{2}}\bigr \rangle = \bigr|{\{{(C\bar{q})}^{1}{(q\bar{q})}^{1}\}^{2}}\bigr \rangle
			\label{spinwfsssqJ2}		
		\end{equation}
	\end{subequations}
	
	Similarly, the recoupling of color wave functions is given by:
	\begin{equation}		
		|{{(Cq)}_{\bar{\textbf{3}}}{(\bar{q}\bar{q})}_{\textbf{3}}}|\rangle=\sqrt{\frac{1}{3}} |\bigr(C\bar{q}\bigr)_{\textbf{1}}\bigr(q\bar{q}\bigr)_{\textbf{1}}\rangle - \sqrt{\frac{2}{3}}|\bigr(C\bar{q}\bigr)_{\textbf{8}}\bigr(q\bar{q}\bigr)_{\textbf{8}}\rangle 
		\label{colorwf1sssq}		
	\end{equation}
	
	where the color representation's dimensions are denoted using subscripts and the total spin is denoted by a superscript. Utilizing the Fierz transformation, $C\bar{q}$ and $q\bar{q}$ are brought together \cite{Ali:2019roi}, and the tetraquark decays into two mesons by the spectator pair technique. The quark bilinears are normalized to unity. The rearrangement decay is described as due to the decay of individual meson pair into lower mass states in the following channels:\begin{itemize}
		
		\item[1.] A spin-0 quarkonium pair in the color-singlet representation decays into two gluons or two photons. The gluons subsequently hadronize into lighter hadrons. This decay process occurs at a rate proportional to \( \alpha_s^2 \). \\
		
		\item[2.] A spin-1 quarkonium pair in the color-singlet representation decays into three gluons or three photons. The gluons again hadronize into lighter hadrons, and the decay rate is of order \( \alpha_s^3 \). \\
		
		\item[3.] Quarkonium pairs in the color-octet representation undergo annihilation: the spin-0 state decays into one gluon, while the spin-1 state decays into two gluons. \\
		
		\item[4.] A spin-0 \( C\bar{q} \) pair in the color-singlet representation can decay into purely leptonic channels of the form \( l^{\pm} + \nu_l \), where \( l \) denotes a lepton (\( e, \mu, \tau \)). \\
		
		\item[5.] The same spin-0 \( C\bar{q} \) state can also undergo rare radiative leptonic decays of the form \( \gamma l^{\pm} + \nu_l \), depending on the meson species. \\
		
		\item[6.] A spin-1 \( C\bar{q} \) pair in the color-singlet representation can decay into dileptonic channels such as \( l^{+} l^{-} \), where \( l = e, \mu, \tau \). \\
		
		\item[7.] A spin-0 \( C\bar{q} \) pair in the color-singlet representation can decay into two body non-leptonic channels using factorization approach. \\
		
	\end{itemize}
	
	The ratio of overlap probabilities of the decaying meson in  tetraquark is proportional to the decay rates: $\varrho = \frac{|\Psi_{\text{Tetraquark}}(0)|^{2}}{|\Psi_{\text{Meson}}(0)|^{2}}$. The individual decay rate is obtained using the formula \cite{Berestetskii:1982qgu}:
	\begin{equation}
		\Gamma(\text{Meson})^{spin}_{color}=||\Psi_{T}(0)|^{2}v\sigma((\text{Meson})^{spin}_{color}\rightarrow f)
	\end{equation}
	where $|\Psi(0)_{T}|^{2},v,$ and $\sigma$ are the overlap probability of the annihilating pair, relative velocity, and the spin-averaged annihilation cross section in the final state $f$, respectively. A more detailed explanation of the working mechanism of the decays described in this section can be found in our previous studies, Refs. \cite{Lodha:2025ffp,Lodha:2024}.The spectator pair $q\bar{q}$ can appear as $\rho$ or $\pi$ on the mass shell. Similarly, the spectator pair $C\bar{q}$ can appear as $D_{0}$ or $D_{1}$ on the mass shell, respectively. The results for calculated the decay widths for the $^{1}S_{0}$, $^{3}S_{1}$ and $^{5}S_{2}$ states are tabulated in table \ref{spectatordecay}, \ref{spectatordecayS1} and \ref{spectatordecayS2} respectively.
	
	\begin{table*}
		\centering
		\caption{Decay width for various decay channels for $T_{^{1}S_{0}}$ in MeV}
		\label{spectatordecay}
		\begin{tabular}{ccc}
			\hline
			Decay Channel    &   Semi-Relativistic   &     Non-Relativistic       \\ \hline
			$D+\text{ 2 gluons}$       & $1.34$ & $2.34$ \\
			
			$\pi + \rho^{0} K^{+}$  & $326.34\times 10^{-15}$ & $461.27\times 10^{-15}$ \\
			
			$D+\text{ 2 photons}$         & $429.26\times10^{-9}$   & $749.02\times10^{-9}$ \\
			
			$\pi + \rho^{+} \phi$ & $2.67\times 10^{-12}$ & $3.85\times 10^{-12}$ \\
			
			$D^{*}+\text{ 3 gluons}$        & $197.12\times10^{-3}$     & $343.96\times10^{-3}$ \\
			
			$\pi + K^{-} \pi^{+}$  & $22.32\times 10^{-12}$ & $31.42\times 10^{-12}$ \\
			
			$D^{*}+\text{ 3 photons}$       & $1.05\times10^{-9}$       & $1.83\times10^{-9}$   \\
			
			$\pi + \pi^{-} \pi^{+}$  & $1.16\times 10^{-12}$ & $1.63\times 10^{-12}$ \\
			
			$D+\text{ 1 gluons}$          & $35.86\times10^{-9}$       & $62.58\times10^{-9}$  \\
			
			$\pi + \pi^{0} \pi^{0}$  & $758.97\times 10^{-15}$ & $1.06\times 10^{-12}$ \\
			
			$D^{*}+\text{ 2 gluons}$        &     $2.66\times10^{-12}$              &     $4.64\times10^{-12}$   \\
			
			 $\pi + K^{-} \rho^{+}$ & $43.48\times 10^{-12}$ & $61.46\times 10^{-12}$ \\
			
			$D^{*}+\mu^{+}\mu^{-}$         & $139.64\times10^{-6}$          & $243.67\times10^{-6}$  \\
			
			$\pi + \eta \bar{K}^{*0}$  & $4.00\times 10^{-12}$ & $5.67\times 10^{-12}$ \\

			$D^{*}+e^{+}e^{-}$           & $139.93\times10^{-6}$          & $244.18\times10^{-6}$\\
			
			$\pi +\omega \phi$  & $378.02\times 10^{-15}$ & $546.45\times 10^{-15}$ \\

			$\pi+ e^{\pm}\nu_{e}$         &$19.35\times10^{-12}$            &$27.03\times10^{-12}$\\
			
			$\pi + \eta^{\prime} \bar{K}^{*0}$  & $16.57\times 10^{-15}$ & $37.84\times 10^{-15}$ \\

			$\pi+ \mu^{\pm}\nu_{\mu}$  						       &$83.11\times10^{-6}$           &$116.13\times10^{-6}$  \\
			      
			$\pi+ \tau^{\pm}\nu_{\tau}$       &$229.51\times10^{-6}$            &$320.68\times10^{-6}$                 \\
			
			$\pi+\gamma l^{\pm}\nu_{l}$      & $13.82\times10^{-9}$   &    $19.31\times10^{-9}$                \\
			
			$\pi + \eta \phi$  & $155.95\times 10^{-15}$ & $222.58\times 10^{-15}$ \\

			$D+ e^{\pm}\nu_{e}$         &$153.99\times10^{-12}$           &$268.70\times10^{-12}$              \\
			
			 $\pi + \rho^{-} \pi^{+}$ & $818.22\times 10^{-15}$ & $1.15\times 10^{-12}$ \\

			$D+ \mu^{\pm}\nu_{\mu}$         &$6.54\times10^{-6}$            &$11.42\times10^{-6}$             \\
			
			$\pi + \pi^{0} \phi$  & $350.49\times 10^{-15}$ & $495.49\times 10^{-15}$ \\
			
			$D+\gamma l^{\pm}\nu_{l}$      &     $1.23\times10^{-6}$  &     $2.15\times10^{-6}$      \\
			
			$\pi + \pi^{-} \rho^{+}$  & $2.40\times 10^{-12}$ & $3.39\times 10^{-12}$ \\
			
			$\pi + \pi^{0} \pi^{+}$  & $2.62\times 10^{-12}$ & $3.68\times 10^{-12}$ \\
			
			$\pi + \rho^{0} \pi^{0}$ & $989.42\times 10^{-15}$ & $1.39\times 10^{-12}$ \\

			$\pi + \pi^{0} K^{+}$  & $502.22\times 10^{-15}$ & $706.30\times 10^{-15}$ \\
			
			$\pi + \eta \omega$  & $723.32\times 10^{-15}$ & $1.02\times 10^{-12}$ \\
			
			$\pi + \eta K^{+}$  & $299.40\times 10^{-15}$ & $421.63\times 10^{-15}$ \\
			
			$\pi + \pi^{0} \omega$  & $815.03\times 10^{-15}$ & $1.15\times 10^{-12}$ \\
			
			$\pi + \eta^{\prime} K^{+}$  & $119.21\times 10^{-15}$ &$ 168.62\times 10^{-15}$\\
			
			$\pi + \eta \eta^{\prime}$  & $20.07\times 10^{-15}$ & $28.46\times 10^{-15}$ \\
			
			$\pi + \eta \pi^{+}$  &$ 312.01\times 10^{-15}$ & $44.00\times 10^{-15}$ \\
			
			$\pi + \eta^{\prime} \pi^{0}$  & $1.98\times 10^{-15}$ & $2.83\times 10^{-15}$ \\

			$\pi + \eta^{\prime} \pi^{+}$  & $3.67\times 10^{-12}$ & $5.18\times 10^{-12}$  \\
			
			$\pi + \eta \pi^{0}$  & $845.07\times 10^{-15}$ & $1.19\times 10^{-12}$ \\

			$\pi + \pi^{+} \rho^{0}$  & $489.55\times 10^{-15}$ & $690.13\times 10^{-15}$ \\
			
			$\pi + \pi^{-} K^{+}$  & $78.51\times 10^{-15}$ & $110.42\times 10^{-15}$ \\

			$\pi + \pi^{+} \phi$  & $2.23\times 10^{-12}$ & $3.16\times 10^{-12}$ \\
			
			$\pi + \eta \eta$ & $837.95\times 10^{-15}$ & $1.18\times 10^{-12}$ \\
			
			$\pi + \pi^{+} \omega$ & $385.37\times 10^{-15}$ & $543.28\times 10^{-15}$ \\
			
			$\pi + K^{-} K^{+}$  & $1.60\times 10^{-12}$ & $2.26\times 10^{-12}$ \\
			
			$\pi + \rho^{0} \rho^{0}$  & $2.24\times 10^{-12}$ & $3.16\times 10^{-12}$ \\
			\hline
			
		\end{tabular}
	\end{table*}

	\begin{table*}
		\centering
		\caption{Decay width for various decay channels for $T_{^{3}S_{1}}$ in MeV}
		\label{spectatordecayS1}
		\begin{tabular}{ccc}
			\hline
			Decay Channel  	& Semi-Relativistic & Non-Relativistic  \\
			\hline
			$D^{*}+\text{ 2 gluons}$ &$2.68$&$4.68$\\
			
			$\rho + \pi^{0} K^{+}$&$ 1.00\times 10^{-12}$ & $ 1.41\times 10^{-12}$ \\
			
			$D^{*}+\text{ 2 photons}$  &$858.52\times10^{-9}$&$1.49\times10^{-6}$\\
			
			$\rho + \eta K^{+}$&$ 598.81\times 10^{-15}$ & $ 843.26\times 10^{-15}$\\
			
			$D+\text{ 3 gluons}$ &$394.24\times10^{-3}$&$687.92\times10^{-3}$\\
			
			$\rho + \eta^{\prime} K^{+}$&$ 238.41\times 10^{-15}$ & $ 337.28\times 10^{-15}$ \\
			
			$D+\text{ 3 photons}$&$2.11\times10^{-9}$&$3.66\times10^{-9}$\\
			
			 $\rho + \pi^{+} \rho^{0}$&$ 979.11\times 10^{-15}$ & $ 1.38\times 10^{-12}$\\
			
			$D^{*}+\text{ 1 gluons} $ & $71.72\times10^{-9}$ &$125.16\times10^{-9}$ \\
			
			$\rho + \pi^{+} \phi$&$ 4.46\times 10^{-12}$ & $ 6.31\times 10^{-12}$ \\
			
			$D+\text{ 2 gluons}$&$5.32\times10^{-12}$&$9.28\times10^{-12}$\\
			
			  $\rho + \pi^{+} \omega$ & $ 770.73\times 10^{-15}$ & $1.08\times 10^{-12}$ \\
			
			$\rho+ e^{\pm}\nu_{e}$ &$38.71\times10^{-12}$&$54.06\times10^{-12}$\\
			
			 $\rho + \rho^{0} K^{+}$&$ 652.68\times 10^{-15}$ & $ 922.54\times 10^{-15}$ \\
			
			$\rho+ \mu^{\pm}\nu_{\mu}$ &$166.22\times10^{-6}$&$232.26\times10^{-6}$ \\
			
			$\rho + \rho^{+} \phi $ & $ 5.34\times 10^{-12}$ & $ 7.71\times 10^{-12}$ \\
			
			$\rho+\gamma  l^{\pm}\nu_{l}$ &$27.64\times10^{-9}$ &$38.62\times10^{-9}$\\
			
			 $\rho + K^{-} \pi^{+}$&$ 44.65\times 10^{-12}$ & $ 62.82\times 10^{-12}$ \\
			
			$\rho + \eta \pi^{+}$&$ 624.01\times 10^{-15}$ & $ 880.01\times 10^{-15}$ \\
			
			 $\rho + \eta^{\prime} \pi^{+}$&$ 7.35\times 10^{-12}$ & $ 10.37 \times 10^{-12}$  \\
			
			$\rho + \pi^{-} \pi^{+}$&$ 2.31\times 10^{-12}$ & $ 3.25\times 10^{-12}$ \\
			
			  $\rho + \pi^{0} \pi^{0}$&$ 1.52\times 10^{-12}$ & $ 2.13\times 10^{-12}$  \\
			
			$\rho + K^{-} K^{+}$&$ 3.21\times 10^{-12}$ & $ 4.51\times 10^{-12}$ \\
			
			  $\rho + \eta \eta $ & $ 1.68\times 10^{-12}$ & $ 2.36\times 10^{-12}$ \\
			
			$\rho + \pi^{-} K^{+}$&$ 157.03\times 10^{-15}$ & $ 220.84\times 10^{-15}$ \\
			
			  $\rho + \eta \pi^{0}$&$ 1.69\times 10^{-12}$ & $ 2.38\times 10^{-12}$  \\
			
			$\rho + \eta^{\prime} \pi^{0}$&$ 3.97\times 10^{-15}$ & $ 5.68\times 10^{-15}$ \\
			
			 $\rho +\omega \phi$&$ 756.04\times 10^{-15}$ & $ 1.09\times 10^{-12}$ \\
			
			$\rho + \eta \eta^{\prime}$&$ 40.14\times 10^{-15}$ & $ 56.93\times 10^{-15}$ \\
			
			$\rho + \pi^{0} \omega$&$ 1.63\times 10^{-12}$ & $ 2.29 \times 10^{-12}$ \\
			
			$\rho + \eta \omega$&$ 1.45\times 10^{-12}$ & $ 2.05\times 10^{-12}$ \\
			
			  $\rho + \rho^{0} \pi^{0}$ & $ 1.97\times 10^{-12}$ & $ 2.79\times 10^{-12}$  \\
			
			$\rho + \pi^{-} \rho^{+}$ &  $ 4.81 \times 10^{-12}$ & $ 6.78\times 10^{-12}$ \\
			
			  $\rho + \pi^{0} \phi$&$ 700.99\times 10^{-15}$ & $ 990.98\times 10^{-15}$  \\
			
			$\rho + \rho^{-} \pi^{+}$ & $ 1.63\times 10^{-12}$ & $ 2.31\times 10^{-12}$ \\
			
			 $\rho + \eta \phi$&$ 311.89\times 10^{-15}$ & $ 445.15\times 10^{-15}$ \\
			
			$\rho + K^{-} \rho^{+}$ & $ 86.97\times 10^{-12}$ & $ 122.93\times 10^{-12}$ \\
			
			$\rho + \eta \bar{K}^{*0}$  & $8.00\times 10^{-12}$ & $ 11.36\times 10^{-12}$  \\
			
			$\rho + \eta^{\prime} \bar{K}^{*0}$   &$ 33.14\times 10^{-15}$ & $ 75.69\times 10^{-15}$ \\
			
			 $\rho + \rho^{0} \rho^{0}$&$  4.49\times 10^{-12}$ & $ 6.33\times 10^{-12}$ \\
			
			$\rho + \pi^{0} \pi^{+}$&$ 5.24\times 10^{-12}$ &  $ 7.37\times 10^{-12}$ \\
			\hline
		\end{tabular}
	\end{table*}
	
	\begin{table*}
		\centering
		\caption{Decay width for various decay channels for $T_{^{5}S_{2}}$ in MeV}
		\label{spectatordecayS2}
		\begin{tabular}{ccc}
			\hline
			Decay Channel  		& Semi-Relativistic & Non-Relativistic  \\
			
			\hline
			$D^{*}+\text{ 3 gluons}$ &$788.48\times10^{-3}$&$1.38$\\
			
			$D+\text{ 1 gluons}$ &$143.44\times10^{-9}$&$250.32\times10^{-9}$\\
			
			$D^{*}+\text{ 3 photons}$ &$4.22\times10^{-9}$&$7.32\times10^{-9}$\\
			
			$D^{*}+\text{ 2 gluons}$&$10.64\times10^{-12}$&$18.56\times10^{-12}$ \\
			\hline& 
		\end{tabular}
	\end{table*}
			
			\section{Regge trajectories}
			\label{sec:4}

			\begin{figure*}[t]
				\centering
				\begin{subfigure}{0.475\textwidth}
					\includegraphics[width=1\linewidth, height=0.3\textheight]{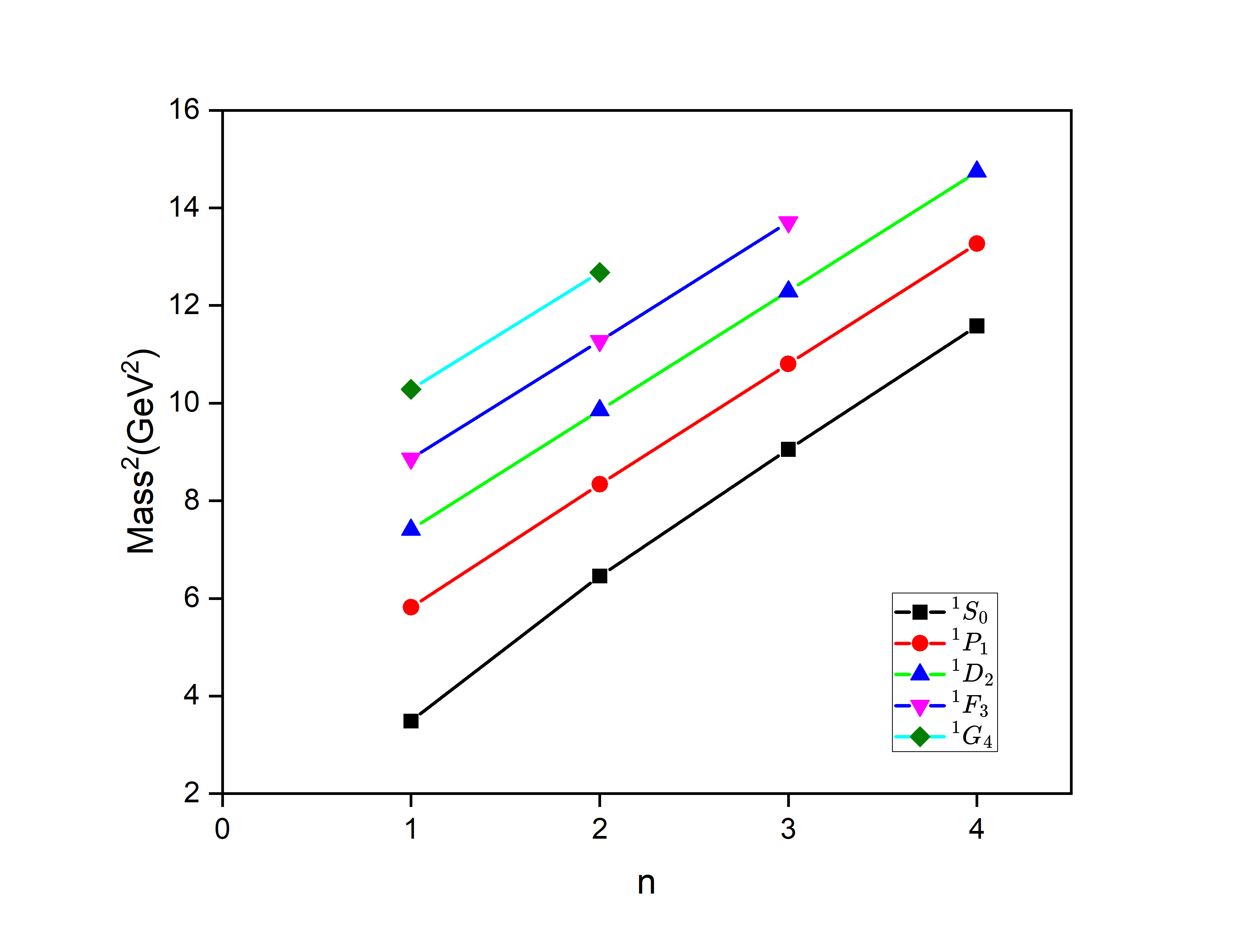}
					\caption{Non-relativistic }
					\label{fig:mesonNR0}
				\end{subfigure}
				\begin{subfigure}{0.475\textwidth}
					\includegraphics[width=1\linewidth, height=0.3\textheight]{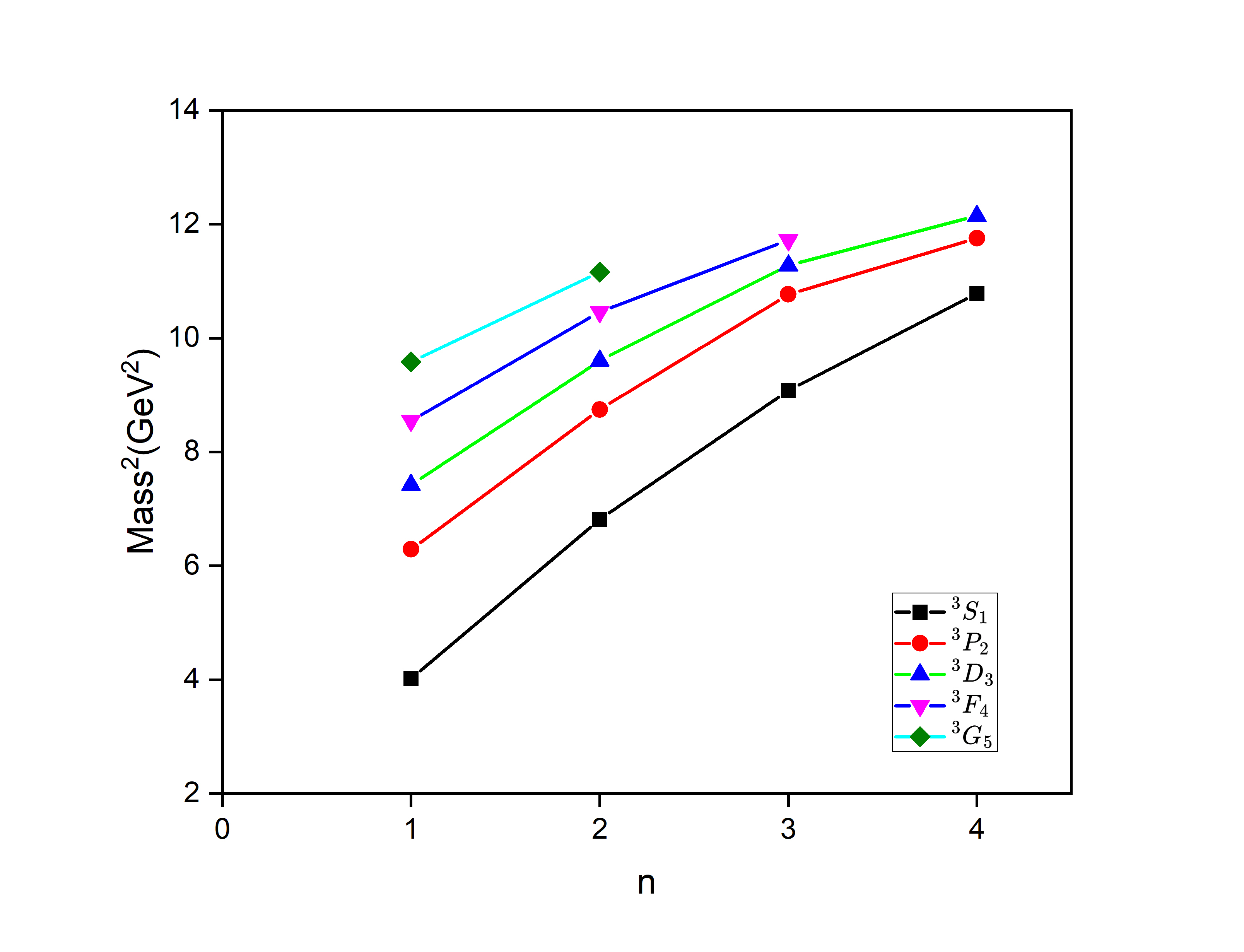}
					\caption{Semi-relativistic}
					\label{fig:mesonSR0}
				\end{subfigure}
				\label{GraphS1}
				\caption[]{Regge trajectory in the $(n, M^{2})$ plane for $D$ meson with Spins S = 0}
			\end{figure*}

			\begin{figure*}[t]
				\centering
				\begin{subfigure}{0.475\textwidth}
					\includegraphics[width=1\linewidth, height=0.3\textheight]{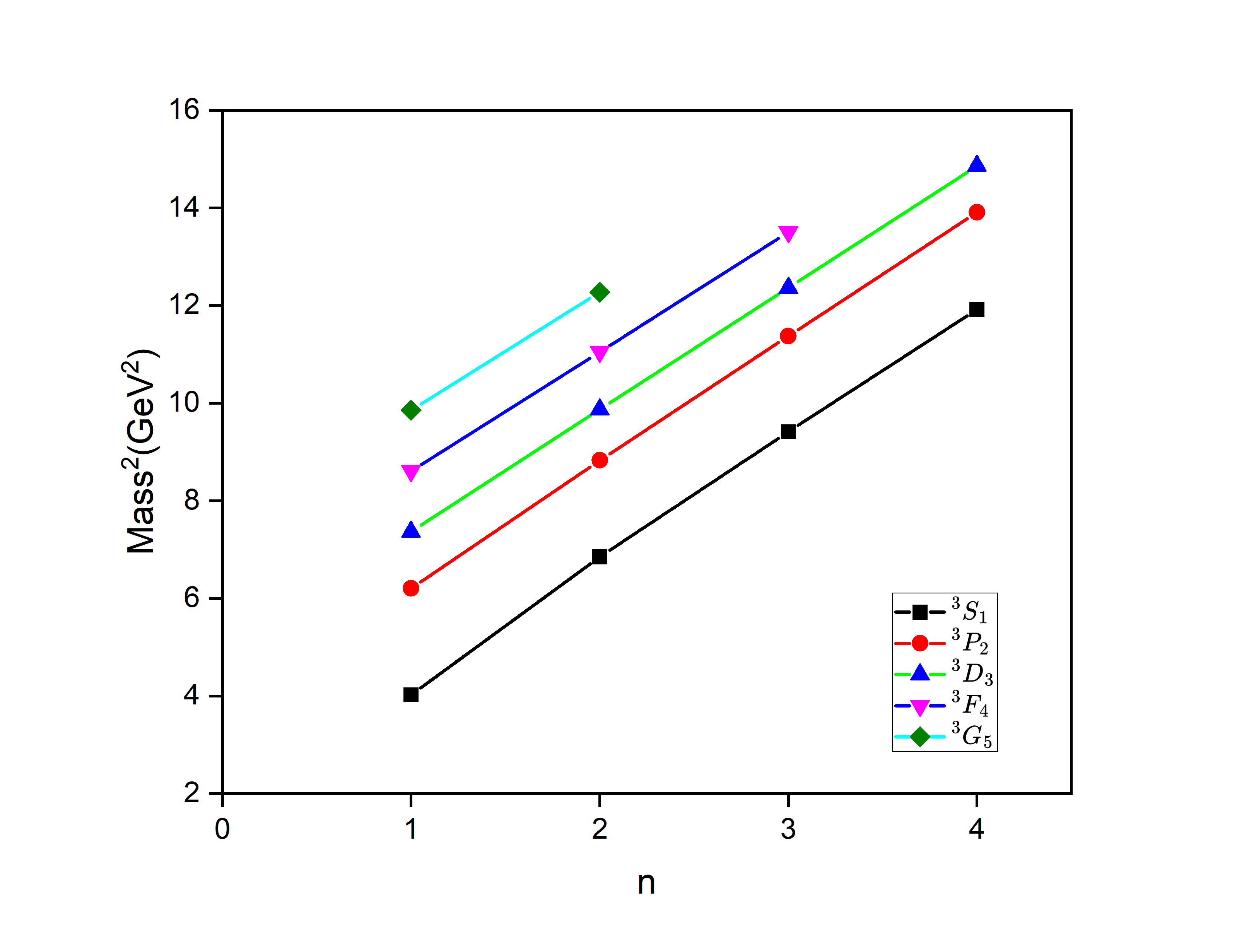}
					\caption{Non-relativistic}
					\label{fig:mesonNR1}
				\end{subfigure}
				\begin{subfigure}{0.475\textwidth}
					\includegraphics[width=1\linewidth, height=0.3\textheight]{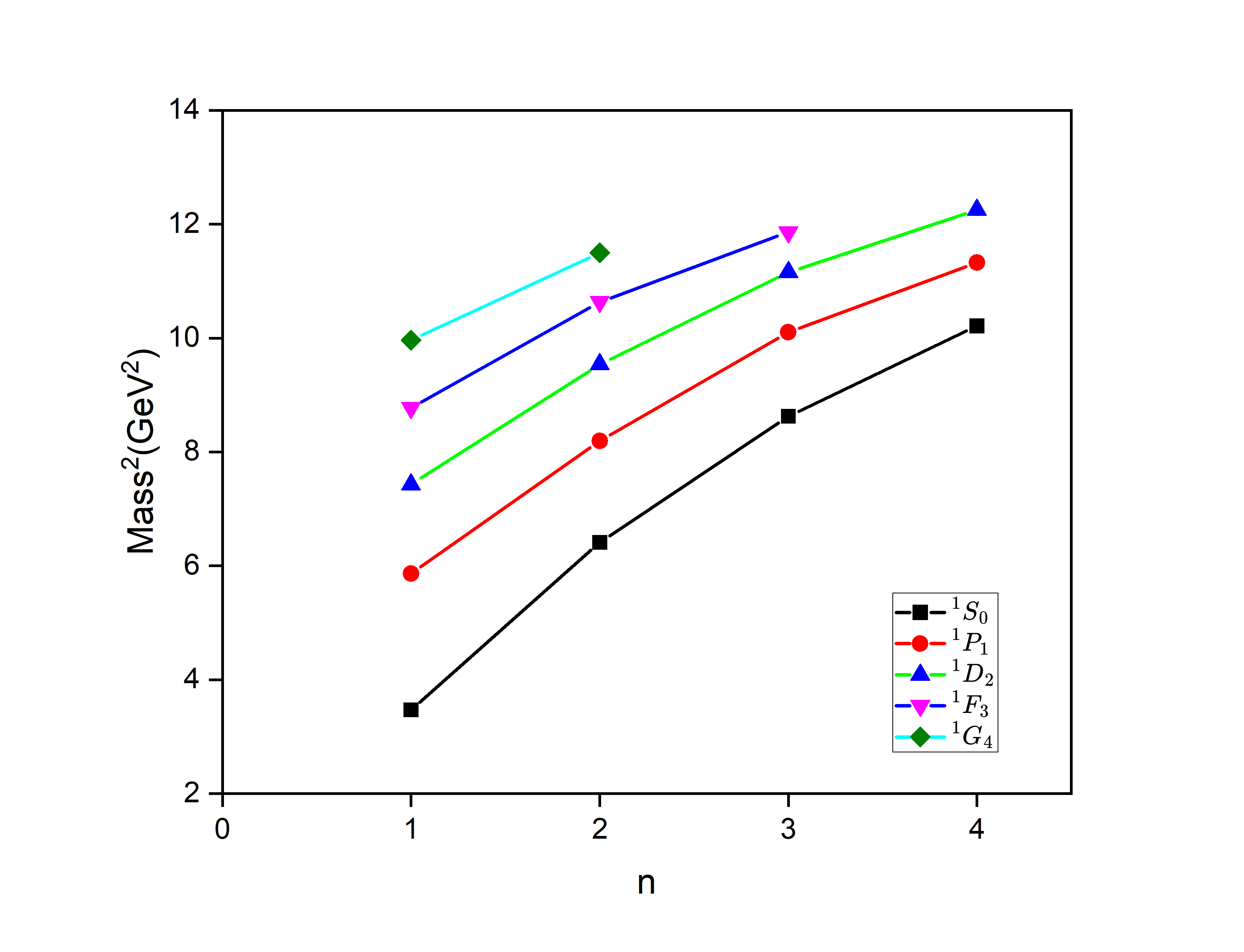}
					\caption{Semi-relativistic}
					\label{fig:mesonSR1}
				\end{subfigure}
				\label{GraphS2}
				\caption[]{Regge trajectory in the $(n, M^{2})$ plane for $D$ meson with Spins S = 1}
			\end{figure*}

			\begin{figure*}[t]
				\centering
				\begin{subfigure}{0.475\textwidth}
					\includegraphics[width=1\linewidth, height=0.3\textheight]{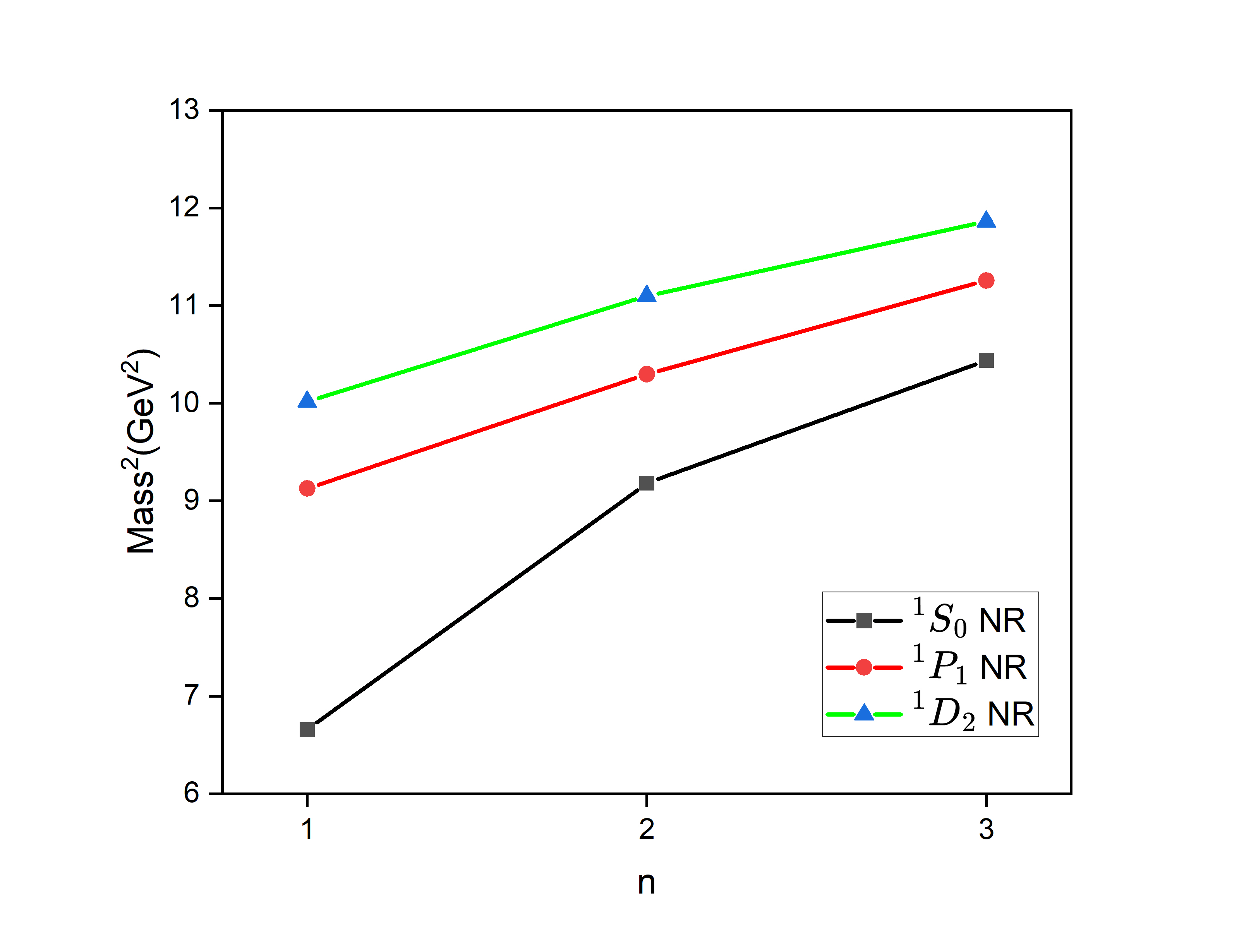}
					\caption{Non-relativistic}
					\label{fig:NRT0}
				\end{subfigure}
				\begin{subfigure}{0.475\textwidth}
					\includegraphics[width=1\linewidth, height=0.3\textheight]{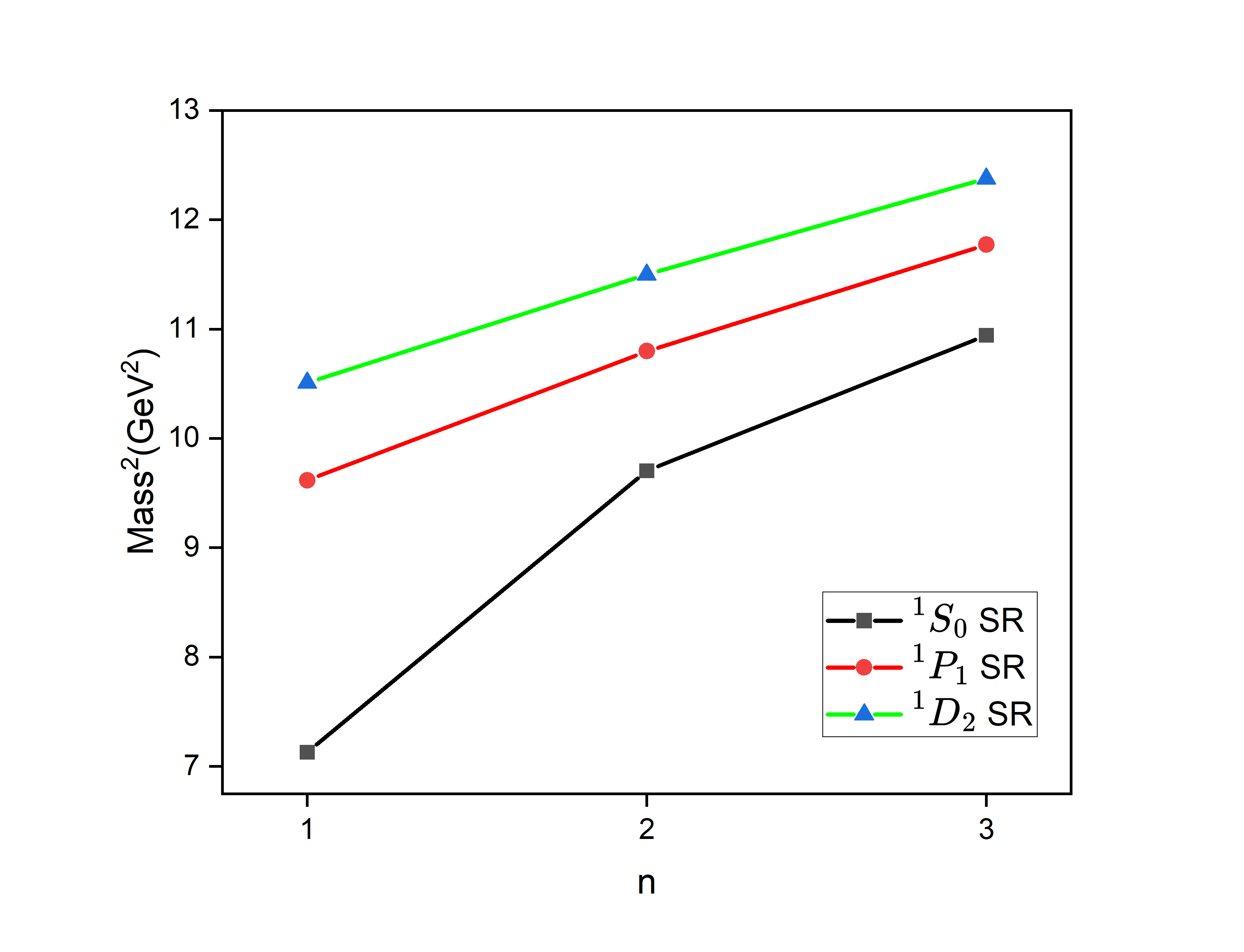}
					\caption{Semi-relativistic}
					\label{fig:SRT0}
				\end{subfigure}
				\label{TetraS0}
				\caption[]{Regge trajectory in the $(n, M^{2})$ plane for $Cq\bar{q}\bar{q}$ tetraquark with Spins S = 0}
			\end{figure*}
			
			\begin{figure*}[t]
				\centering
				\begin{subfigure}{0.475\textwidth}
					\includegraphics[width=1\linewidth, height=0.3\textheight]{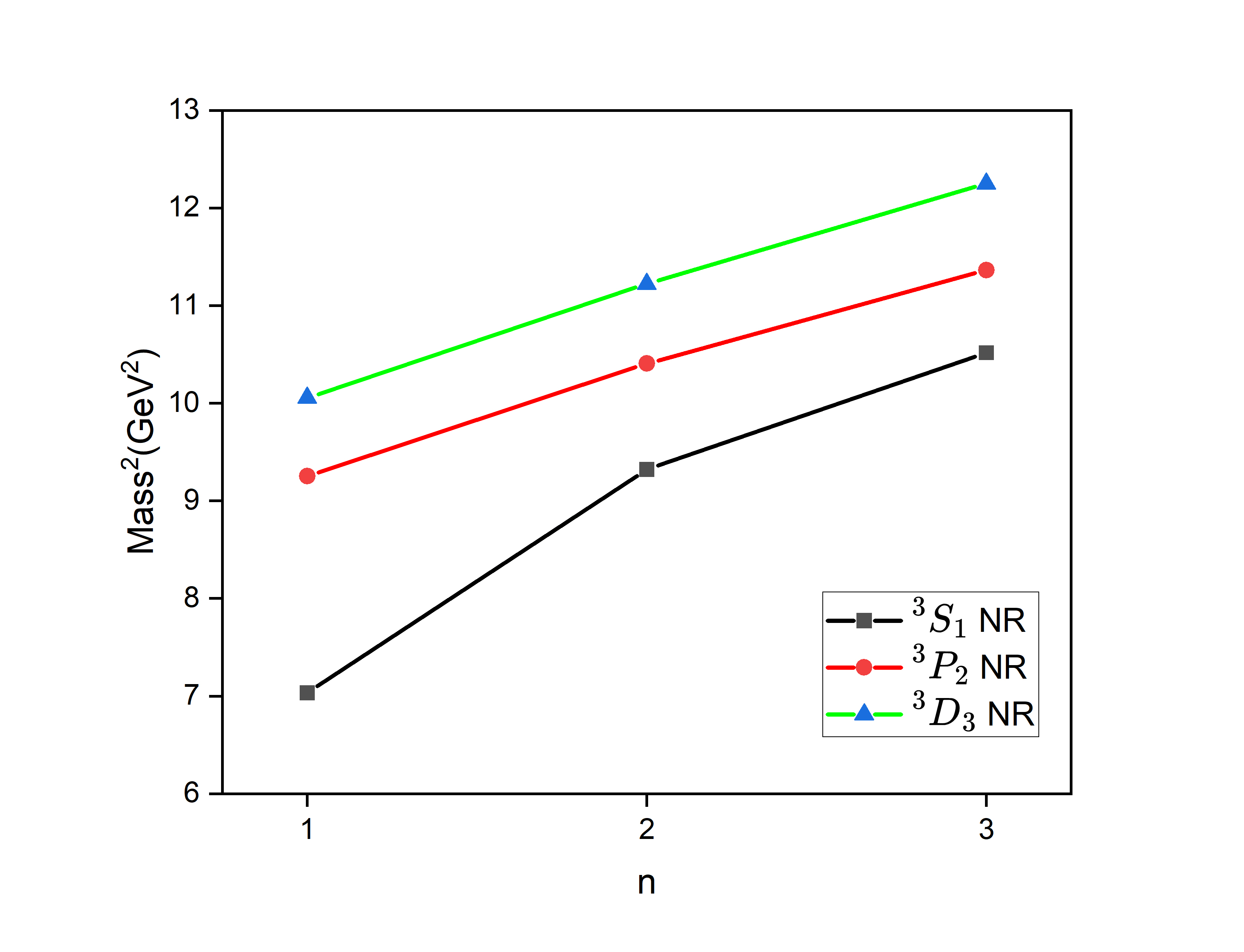}
					\caption{Non-relativistic}
					\label{fig:NRT1}
				\end{subfigure}
				\begin{subfigure}{0.475\textwidth}
					\includegraphics[width=1\linewidth, height=0.3\textheight]{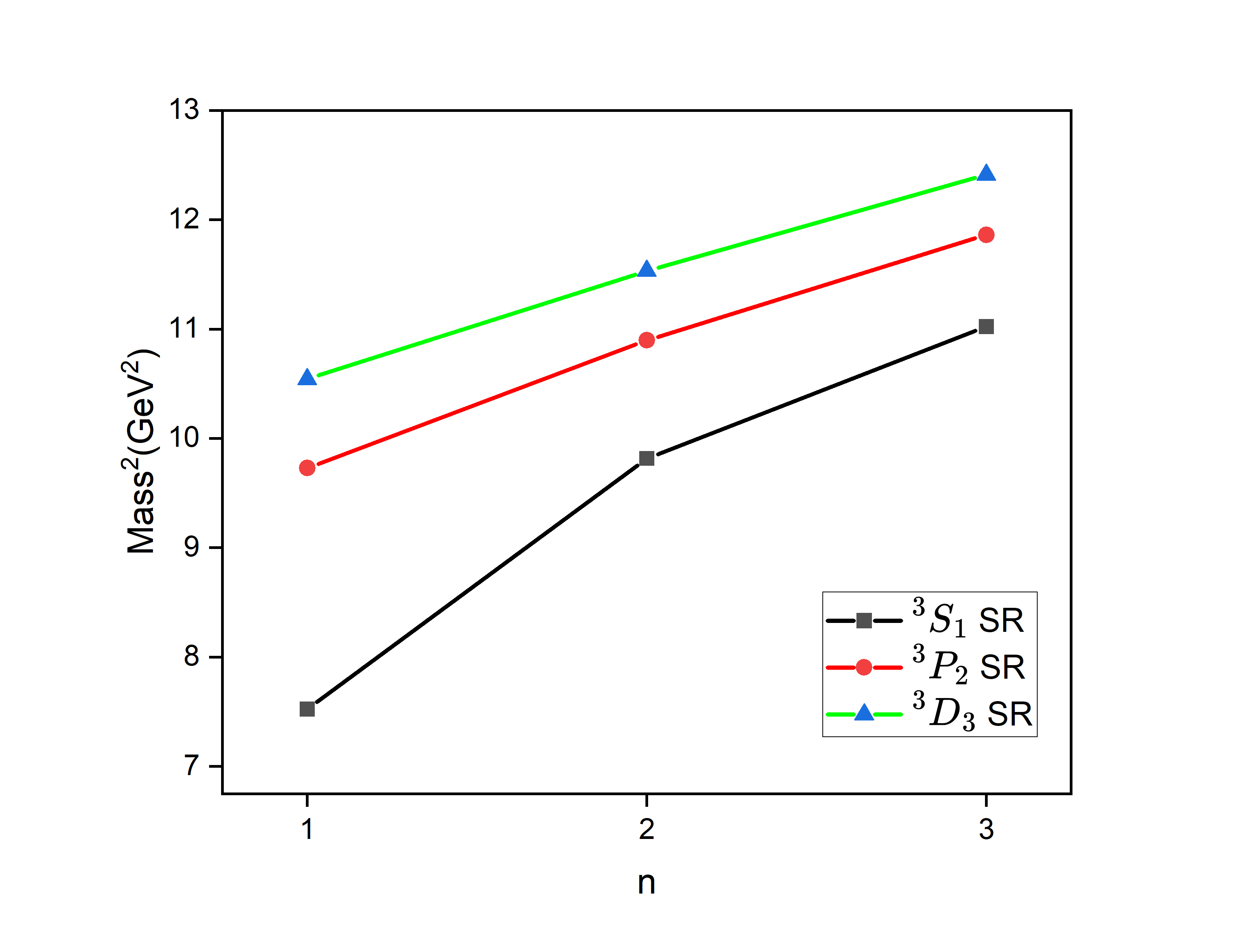}
					\caption{Semi-relativistic}
					\label{fig:SRT1}
				\end{subfigure}
				\label{TetraS1}
				\caption[]{Regge trajectory in the $(n, M^{2})$ plane for $Cq\bar{q}\bar{q}$ tetraquark with Spins S = 1}
			\end{figure*}
			
			\begin{figure*}[t]
				\centering
				\begin{subfigure}{0.475\textwidth}
					\includegraphics[width=1\linewidth, height=0.3\textheight]{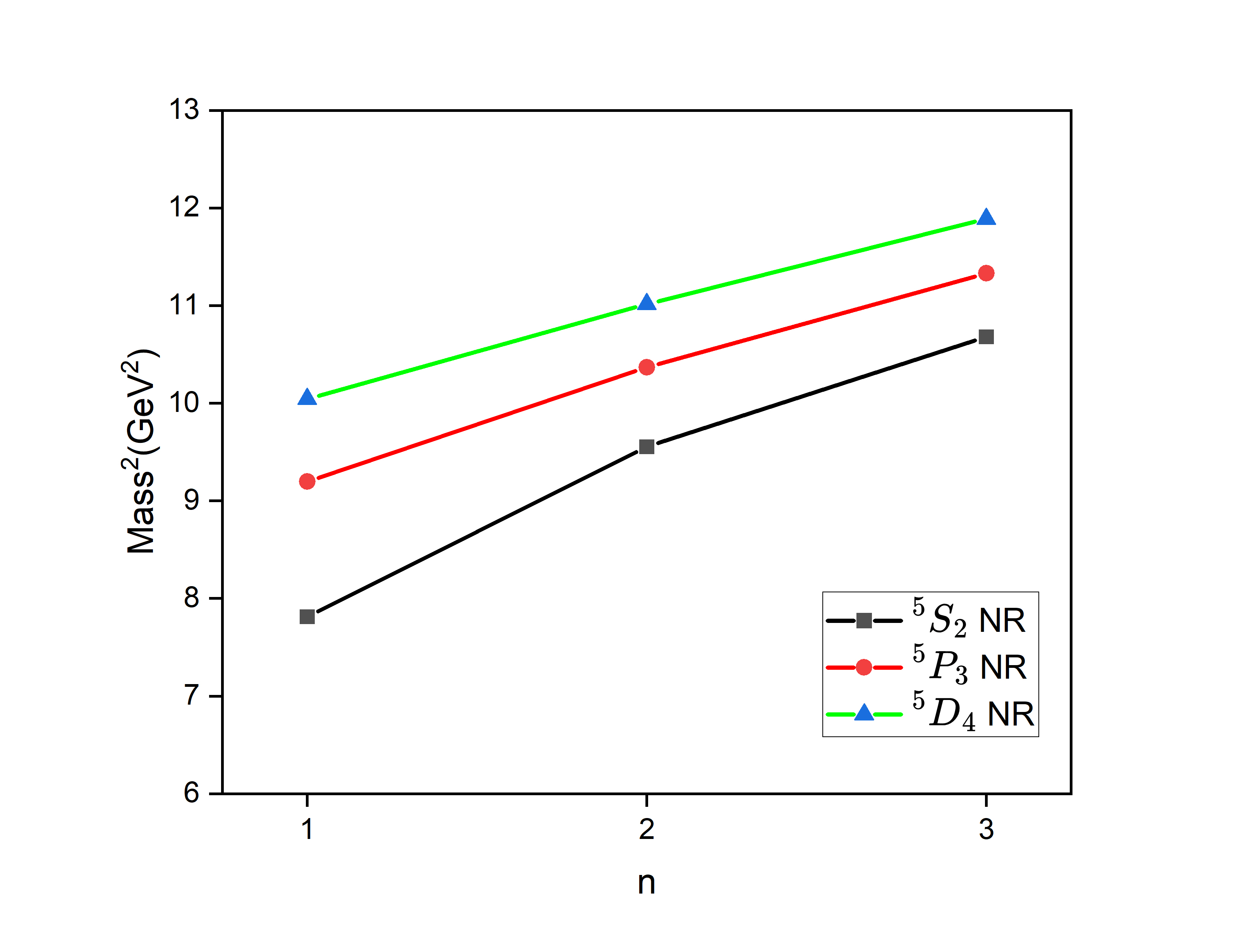}
					\caption{Non-relativistic}
					\label{fig:NRT2}
				\end{subfigure}
				\begin{subfigure}{0.475\textwidth}
					\includegraphics[width=1\linewidth, height=0.3\textheight]{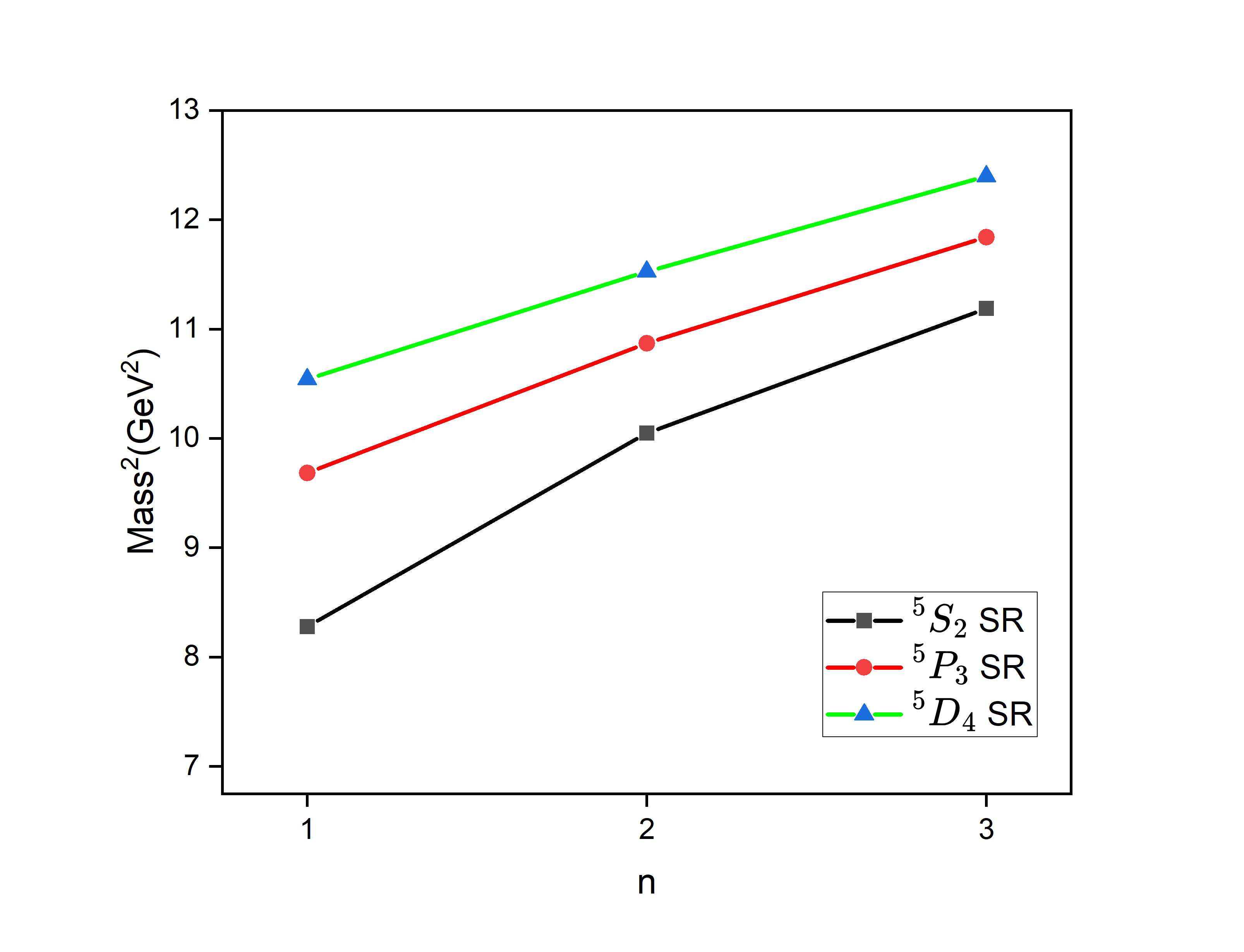}
					\caption{Semi-relativistic}
					\label{fig:SRT2}
				\end{subfigure}
				\label{TetraS2}
				\caption[]{Regge trajectory in the $(n, M^{2})$ plane for $Cq\bar{q}\bar{q}$ tetraquark with Spins S = 2}
			\end{figure*}
			
			Regge trajectories provide a powerful framework for organizing hadron spectra by relating a particle’s spin \(J\) to its squared mass \(M^2\).  Originally developed to describe high-energy scattering amplitudes, Regge theory has become indispensable for classifying mesons, baryons, and exotic states.  In its simplest form, a trajectory is written as
			$J = \alpha'\,M^2 + \alpha_0\,,$ where \(\alpha'\) (the slope) and \(\alpha_0\) (the intercept) encode the confinement dynamics. For light mesons—composed of \(u\), \(d\), and \(s\) quarks—the trajectories are nearly linear but exhibit slight convexity, a consequence of relativistic quark motion and the non-linear nature of the strong interaction at low energies.  Here the gluonic field dominates, producing an almost uniform string tension. By contrast, heavy-light mesons have larger mass suppressesing relativistic effects and causes the trajectories to bend downward (concave behavior).  In these systems, the energy needed to increase \(J\) grows more slowly than \(M^2\), reflecting a more rigid, rotor-like motion; empirically one finds reduced slopes. Exotic tetraquarks, modeled here in the diquark–antidiquark picture, show even richer behavior.  Their multi-body dynamics and internal excitation modes (\(\rho\)- and \(\lambda\)-modes) lead to strongly nonlinear, concave Regge lines in both the \((J,M^2)\) and \((n,M^2)\) planes, where \(n\) is the radial quantum number.  
			
			In this work, we construct Regge trajectories for the calculated mass spectra of \(D\) mesons and \(Cq\bar{q}\bar{q}\) tetraquarks.   The Regge trajectories in the $(n, M^{2})$ plane are drawn for $D$ meson for unnatural and natural parity  as shown in Figs. \ref{fig:mesonNR0}, \ref{fig:mesonSR0}, \ref{fig:mesonNR1} and  \ref{fig:mesonSR1}. Similarly, The Regge trajectories in the $(n, M^{2})$ plane are drawn for $Cq\bar{q}\bar{q}$ tetraquark for Spin $S=0,1$ and 2 are shown in figure \ref{fig:NRT0}, \ref{fig:SRT0}, \ref{fig:NRT1}, \ref{fig:SRT1}, \ref{fig:NRT2} and \ref{fig:SRT2}, respectively.

			\section{Results and Discussion}
			\label{sec:5}
			In the present work, we have calculated the mass spectra of the   $Cq\bar{q}\bar{q}$ tetraquarks by the parameters obtained from calculating the mass spectra of $D$ meson. The mass spectra of $D$ meson, $Cq$ diquark and $Cq\bar{q}\bar{q}$ tetraquark are tabulated in table \ref{mass_meson}, \ref{diquark} and \ref{mass_tetra}, respectively. These masses are calculated for the non-relativistic and semi-relativistic framework, where the semi-relativistic framework also incorporates correction to the kinetic energy upto \(\mathcal{O}(p^{6})\). The decay properties of $D$ meson by factorization approach is calculated and tabulated in table \ref{Branch} which are furthur used to calculate the decay width of the tetraquark for $^{1}S_{0},^{3}S_{1}$ and $^{5}S_{2}$ state, which are shown in table \ref{spectatordecay}, \ref{spectatordecayS1} and \ref{spectatordecayS2} respectively. Since no PDG data for strange tetraquarks is available, the two-meson threshold limit is used as a reference for comparison, which has been tabulated in table \ref{twomesonthreshold}.
			
			\subsection{The $D$ Meson}
			
			In the present work, the mass spectra of $D$ mesons have been calculated within two distinct frameworks: the non-relativistic formalism in the energy range of 1.8 to 3.5~GeV, and the semi-relativistic formalism in the range of 1.8 to 3.3~GeV. The obtained masses are tabulated in Table~\ref{mass_meson}, alongside values reported in other theoretical studies. Employing the factorization approach based on the effective weak Hamiltonian, we also compute the branching fractions for several non-leptonic decay modes of $D$ mesons. These results, presented in Table~\ref{Branch}, are compared with values from other theoretical models and experimental data reported by the Particle Data Group (PDG). Furthermore, Regge trajectories are constructed in the $(J, M^2)$ plane for natural and unnatural parity states, as shown in Figures~\ref{fig:mesonNR0}, \ref{fig:mesonSR0}, \ref{fig:mesonNR1}, and \ref{fig:mesonSR1}. In the non-relativistic case, the trajectories appear nearly linear, while the semi-relativistic results exhibit slight concavity.
			
			Several experimentally observed resonances align closely with our theoretical predictions. In the $S$-wave sector, six resonances—$D^{0}$, $D^{\pm}$, $D^{*}(2007)^{0}$, $D^{*}(2010)^{\pm}$, $D_0(2550)^0$, and $D_1^{*}(2600)^0$—show good agreement. The $D^{0}$ meson has been observed with an average mass of $1864.84 \pm 0.05$~MeV and a mean lifetime of $(4.103 \pm 0.010)\times 10^{-13}$~s~\cite{Goldhaber:1976xn}. The $D^{\pm}$ meson has a measured mass of $1869.66 \pm 0.05$~MeV and a lifetime of $(1.033 \pm 0.005)\times 10^{-12}$~s. Both have isospin-spin-parity $I(J^P) = \frac{1}{2}(0^-)$, matching the $1^{1}S_0$ state in our calculations, with theoretical masses of $1867.49 \pm 0.33$~MeV (non-relativistic) and $1863.67 \pm 0.25$~MeV (semi-relativistic). The $D^{*}(2007)^0$ and $D^{*}(2010)^{\pm}$ resonances, with $I(J^P) = \frac{1}{2}(1^-)$, correspond well to the $1^{3}S_1$ state, whose calculated masses are $2006.72 \pm 0.46$~MeV (NR) and $2005.65 \pm 0.47$~MeV (SR). The $D_0(2550)^0$ resonance, with $I(J^P) = \frac{1}{2}(0^-)$, has been observed with an average mass of $2549 \pm 19$~MeV and width of $165 \pm 25$~MeV in processes such as $B^{-} \to D^{*+}\pi^{-}\pi^{-}$, $pp \to D^{*+}\pi^{-}X$, and $e^{+}e^{-} \to D^{*+}\pi^{-}X$~\cite{LHCb:2013jjb,BaBar:2010zpy,LHCb:2019juy}. This aligns well with the $2^{1}S_0$ state in our model. Similarly, the $D_1^{*}(2600)^0$ resonance, observed in multiple decay modes~\cite{LHCb:2013jjb,BaBar:2010zpy,LHCb:2019juy,LHCb:2016lxy}, is well represented by the $2^{3}S_1$ state with $I(J^P) = \frac{1}{2}(1^-)$.
			
			In the $P$-wave sector, four resonances show promising agreement: $D^{*}(2300)$, $D_1(2420)$, $D_1(2430)^0$, and $D_2^{*}(2460)$. The $D^{*}(2300)$ resonance, with $I(J^P) = \frac{1}{2}(0^+)$ and mass $2343 \pm 10$~MeV, has been observed in various processes~\cite{LHCb:2015tsv,LHCb:2015klp,BaBar:2009pnd,Belle:2003nsh}, and corresponds well to the $1^{3}P_0$ state of our model. The $D_1(2420)$ and $D_1(2430)^0$ resonances, both with $I(J^P) = \frac{1}{2}(1^+)$ and masses $2422.1 \pm 0.6$~MeV and $2412 \pm 9$~MeV respectively, are candidates for the $1^{1}P_1$ or $1^{3}P_1$ states~\cite{Belle:2003nsh,LHCb:2019juy,BESIII:2019phe,Belle:2004bvv}. The $D_2^{*}(2460)$ state, with $I(J^P) = \frac{1}{2}(2^+)$, mass $2461.1 \pm 0.8$~MeV, and decay width $47.3 \pm 0.8$~MeV~\cite{LHCb:2015klp,LHCb:2015eqv,LHCb:2016lxy,LHCb:2015tsv}, aligns well with our predicted $1^{3}P_2$ state.
			
			In the $D$-wave sector, three resonances can be confidently associated: $D_2(2740)^0$, $D_3(2750)$, and $D_1^{*}(2760)^0$. The $D_2(2740)^0$ resonance, with $I(J^P) = \frac{1}{2}(2^-)$ and mass $2747 \pm 6$~MeV~\cite{LHCb:2013jjb,LHCb:2019juy}, is consistent with either the $1^{1}D_2$ or $1^{3}D_2$ states of our spectrum. The $D_3(2750)$, observed with mass $2763.1 \pm 3.2$~MeV and decay width $66 \pm 5$~MeV~\cite{LHCb:2016lxy,LHCb:2015klp,LHCb:2013jjb,BaBar:2010zpy}, corresponds to the $1^{3}D_3$ state. Lastly, the $D_1^{*}(2760)^0$ resonance, observed in $B^- \to D^+K^-\pi^-$ with a mass of $2781 \pm 18 \pm 13$~MeV and width of $180 \pm 40$~MeV~\cite{LHCb:2015eqv}, is a strong candidate for the $1^{3}D_1$ state with $I(J^P) = \frac{1}{2}(1^-)$. Among the discussed resonances, $D_0(2550)^0$, $D_1^{*}(2600)^0$, $D_2(2740)^0$, and $D_1^{*}(2760)^0$ are still considered unconfirmed or not well established according to the PDG.
			
			\subsection{The $Cq\bar{q}\bar{q}$ Tetraquark}
			
			The present study estimates the mass of the $Cq\bar{q}\bar{q}$ tetraquark in the $\bar{\mathbf{3}}$–$\mathbf{3}$ color configuration to lie between 2.5~GeV and 3.5~GeV using the non-relativistic formalism, and between 2.6~GeV and 3.5~GeV using the semi-relativistic formalism. These mass ranges make such states ideal candidates for exploration at experimental facilities like LHCb, Belle II, BESIII, and \textsc{P}ANDA. The decay widths for various $S$-wave tetraquark states are summarized in Tables~\ref{spectatordecay}, \ref{spectatordecayS1}, and \ref{spectatordecayS2}.
			
			For the $1^{1}S_{0}$ state, the dominant decay mode is $D + 2$ gluons, with decay widths of 1.34~MeV (semi-relativistic) and 2.34~MeV (non-relativistic). The next most significant decay channel is $D^{*} + 3$ gluons, with decay widths of 0.197~MeV and 0.343~MeV, respectively. In the case of the $1^{3}S_{1}$ state, the primary decay channel is $D^{*} + 2$ gluons, yielding widths of 2.68~MeV and 4.68~MeV for the semi-relativistic and non-relativistic formalisms, respectively. The secondary channel, $D + 3$ gluons, has corresponding widths of 0.39~MeV and 0.68~MeV. For the $^{5}S_{2}$ state, the decay mode $D^{*} + 3$ gluons is dominant, with decay widths of 0.788~MeV (semi-relativistic) and 1.38~MeV (non-relativistic).
			
			Two experimentally observed resonances may be interpreted as candidates for the singly charmed tetraquark configurations studied here. The first, $D^{*}(2640)^{\pm}$, was observed by the DELPHI collaboration in $66 \pm 14$ events, with a reported mass of $2637 \pm 2 \pm 6$~MeV and an undefined $J^{P}$ quantum number~\cite{DELPHI:1998oyl}. This state appeared in the process $e^{+}e^{-} \rightarrow D^{*+}\pi^{+}\pi^{-}X$ with a decay width measured to be $< 15$~MeV. Although not well established in the PDG listings, this resonance aligns well with the $1^{1}S_{0}$ tetraquark state, whose mass is predicted to be $2580.65 \pm 0.64$~MeV (non-relativistic) and $2689.92 \pm 0.63$~MeV (semi-relativistic).
			
			The second candidate, $D(3000)^{0}$, has been observed in several processes: $B^{-} \rightarrow D^{+}\pi^{-}\pi^{-}$, $pp \rightarrow D^{*+}\pi^{-}X$, and $pp \rightarrow D^{+}\pi^{-}X$, with reported masses of $3214 \pm 29 \pm 49$~MeV, $2971.8 \pm 8.7$~MeV, and $3008.1 \pm 4.0$~MeV, respectively~\cite{LHCb:2016lxy, LHCb:2013jjb}. This state also has an undefined $J^{P}$ assignment and a relatively broad decay width of $190 \pm 80$~MeV. As a result, multiple theoretical interpretations are possible, and a definitive spin-parity measurement would greatly refine the classification of this resonance.
			
			\section*{Conclusion}
			
			In this work, the mass spectra of the $D$ meson and singly charmed tetraquarks have been systematically computed using both non-relativistic and semi-relativistic formalisms. The resulting states have been assigned appropriate $J^{PC}$ quantum numbers based on their internal configurations and dynamical behavior. Additionally, decay widths for various accessible channels have been evaluated: the factorization approach was employed for mesonic decays, while the quark rearrangement model was applied to tetraquark decays. Regge trajectories were constructed for both $D$ mesons and tetraquarks, offering insights into the linearity and clustering behavior of states with different spin-parity assignments.
			
			A total of thirteen experimentally observed resonances have been analyzed as potential assignments to $D$ meson states: $D^{\pm}$, $D^{0}$, $D(2007)^{0}$, $D(2010)^{\pm}$, $D_0(2300)$, $D_1(2420)$, $D_1(2430)^{0}$, $D_2(2460)$, $D_0(2550)^{0}$, $D_1(2600)^{0}$, $D_2(2740)^{0}$, $D_3(2750)$, and $D_1(2760)^{0}$. In parallel, two experimentally observed resonances—$D^{*}(2640)^{\pm}$ and $D(3000)^{0}$—have been explored as potential candidates for the $Cq\bar{q}\bar{q}$ tetraquark configurations studied in this framework.
			
			Future directions will involve extending this formalism to tetraquark states with varying quark content, including double charm and hidden charm sectors. Theoretical predictions will be further refined through the evaluation of additional observables, such as magnetic moments, radiative transitions, and electromagnetic form factors. These features may yield a more comprehensive understanding of the internal structure and dynamics of exotic hadrons.
			
			The results of this study offer useful theoretical guidance for ongoing and upcoming experimental efforts at PANDA, J-PARC, Belle II, BESIII, and LHCb. Continued collaboration between theoretical modeling and experimental validation will be crucial in advancing our understanding of QCD in the non-perturbative regime and in uncovering the full spectrum of hadronic matter.
			
			\section{Data Availability Statement} The datasets generated during and/or analysed during the current study are available from the corresponding author on reasonable request.

			%


\begin{thebibliography}{}
				
				\bibitem{Gell-Mann:1964ewy} 
				M.~Gell-Mann,
				Phys. Lett. \textbf{8}, 214-215 (1964)
				
				\bibitem{Berwein:2024ztx}
				M.~Berwein, N.~Brambilla, A.~Mohapatra and A.~Vairo,
				Phys. Rev. D \textbf{110}, no.9, 094040 (2024)
				
				\bibitem{Belle:2003nnu}
				S.~K.~Choi \textit{et al.} [Belle],
				Phys. Rev. Lett. \textbf{91}, 262001 (2003)
				
				\bibitem{Belle:2014nuw}
				K.~Chilikin \textit{et al.} [Belle],
				Phys. Rev. D \textbf{90}, no.11, 112009 (2014)
				
				\bibitem{Belle:2007umv}
				X.~L.~Wang \textit{et al.} [Belle],
				Phys. Rev. Lett. \textbf{99}, 142002 (2007)
				
				\bibitem{CDF:2009jgo}
				T.~Aaltonen \textit{et al.} [CDF],
				Phys. Rev. Lett. \textbf{102}, 242002 (2009)
				
				\bibitem{D0:2016mwd}
				V.~M.~Abazov \textit{et al.} [D0],
				Phys. Rev. Lett. \textbf{117}, no.2, 022003 (2016)
				
				\bibitem{BESIII:2013ris}
				M.~Ablikim \textit{et al.} [BESIII],
				Phys. Rev. Lett. \textbf{110}, 252001 (2013)
				
				\bibitem{LHCb:2020bwg}
				R.~Aaij \textit{et al.} [LHCb],
				Sci. Bull. \textbf{65}, no.23, 1983-1993 (2020)
				
				\bibitem{Goldhaber:1976xn}
				G.~Goldhaber, F.~Pierre, G.~S.~Abrams, M.~S.~Alam, A.~Boyarski, M.~Breidenbach, W.~C.~Carithers, W.~Chinowsky, S.~Cooper and R.~DeVoe, \textit{et al.}
				Phys. Rev. Lett. \textbf{37}, 255-259 (1976)
				
				\bibitem{LHCb:2019hro}
				R.~Aaij \textit{et al.} [LHCb],
				Phys. Rev. Lett. \textbf{122}, no.21, 211803 (2019)
				
				\bibitem{LHCb:2021ykz}
				R.~Aaij \textit{et al.} [LHCb],
				Phys. Rev. Lett. \textbf{127}, no.11, 111801 (2021)
				
				\bibitem{Bigi:2000wn}
				I.~I.~Y.~Bigi and N.~G.~Uraltsev,
				Nucl. Phys. B \textbf{592}, 92-106 (2001)
				
				
				\bibitem{Owen:2024qij}
				P.~Owen and N.~Serra,
				Eur. Phys. J. ST \textbf{233}, no.2, 225-240 (2024)
				
				\bibitem{Belle-II:2018jsg}
				E.~Kou \textit{et al.} [Belle-II],
				PTEP \textbf{2019}, no.12, 123C01 (2019)
				[erratum: PTEP \textbf{2020}, no.2, 029201 (2020)]
				
				\bibitem{Gotzen:2024agc}
				K.~G\"otzen \textit{et al.} [PANDA],
				Nuovo Cim. C \textbf{47}, no.4, 179 (2024)
				
				\bibitem{HeavyFlavorAveragingGroupHFLAV:2024ctg}
				S.~Banerjee \textit{et al.} [Heavy Flavor Averaging Group (HFLAV)],
				[arXiv:2411.18639 [hep-ex]].
				
				\bibitem{LHCb:2022cak}
				R.~Aaij \textit{et al.} [LHCb],
				Phys. Rev. D \textbf{108}, no.5, 052005 (2023)
				
				
				\bibitem{Belle:2024vho}
				I.~Adachi \textit{et al.} [Belle and Belle-II],
				Phys. Rev. D \textbf{111}, no.1, 012015 (2025)
				
				\bibitem{Buccella:1994nf}
				F.~Buccella, M.~Lusignoli, G.~Miele, A.~Pugliese and P.~Santorelli,
				Phys. Rev. D \textbf{51}, 3478-3486 (1995)
				
				\bibitem{Falk:2001hx}
				A.~F.~Falk, Y.~Grossman, Z.~Ligeti and A.~A.~Petrov,
				Phys. Rev. D \textbf{65}, 054034 (2002)
				
				\bibitem{Grossman:2006jg}
				Y.~Grossman, A.~L.~Kagan and Y.~Nir,
				Phys. Rev. D \textbf{75}, 036008 (2007)
				
				\bibitem{Golowich:2007ka}
				E.~Golowich, J.~Hewett, S.~Pakvasa and A.~A.~Petrov,
				Phys. Rev. D \textbf{76}, 095009 (2007)
				
				\bibitem{Isidori:2011qw}
				G.~Isidori, J.~F.~Kamenik, Z.~Ligeti and G.~Perez,
				Phys. Lett. B \textbf{711}, 46-51 (2012)
				
				
				\bibitem{Patel:2022hhl}
				V.~Patel, R.~Chaturvedi and A.~K.~Rai,
				[arXiv:2201.01120 [hep-ph]].
				
				\bibitem{Devlani:2013kta}
				N.~Devlani and A.~K.~Rai,
				Int. J. Theor. Phys. \textbf{52}, 2196-2208 (2013)
				
				
				\bibitem{Kher:2017wsq}
				V.~Kher, N.~Devlani and A.~K.~Rai,
				Chin. Phys. C \textbf{41}, no.7, 073101 (2017)
				
				
				\bibitem{Oudichhya:2024ikt}
				J.~Oudichhya and A.~K.~Rai,
				Int. J. Mod. Phys. A \textbf{39}, no.28, 2443004 (2024)
				
				\bibitem{Godfrey:1985xj}
				S.~Godfrey and N.~Isgur,
				Phys. Rev. D \textbf{32}, 189-231 (1985)
				
				\bibitem{Ni:2021pce}
				R.~H.~Ni, Q.~Li and X.~H.~Zhong,
				Phys. Rev. D \textbf{105}, no.5, 056006 (2022)
				
				\bibitem{Moir:2016srx}
				G.~Moir, M.~Peardon, S.~M.~Ryan, C.~E.~Thomas and D.~J.~Wilson,
				JHEP \textbf{10}, 011 (2016)
				
				
				
				
				
				
				\bibitem{Cheng:2010vk}
				H.~Y.~Cheng and C.~W.~Chiang,
				Phys. Rev. D \textbf{81}, 074031 (2010)
				
				\bibitem{Korner:1992wi}
				J.~G.~Korner and M.~Kramer,
				Z. Phys. C \textbf{55}, 659-670 (1992)
				
				\bibitem{BaBar:2003oey}
				B.~Aubert \textit{et al.} [BaBar],
				Phys. Rev. Lett. \textbf{90}, 242001 (2003)
				
				\bibitem{CLEO:2003ggt}
				D.~Besson \textit{et al.} [CLEO],
				Phys. Rev. D \textbf{68}, 032002 (2003)
				[erratum: Phys. Rev. D \textbf{75}, 119908 (2007)]
				
				\bibitem{Belle:2003guh}
				P.~Krokovny \textit{et al.} [Belle],
				Phys. Rev. Lett. \textbf{91}, 262002 (2003)
				
				\bibitem{ParticleDataGroup:2024cfk}
				S.~Navas \textit{et al.} [Particle Data Group],
				Phys. Rev. D \textbf{110}, no.3, 030001 (2024)
				
				\bibitem{vanBeveren:2006st}
				E.~van Beveren and G.~Rupp,
				Phys. Rev. Lett. \textbf{97}, 202001 (2006)
				
				\bibitem{BaBar:2006gme}
				B.~Aubert \textit{et al.} [BaBar],
				Phys. Rev. Lett. \textbf{97}, 222001 (2006)
				
				\bibitem{LHCb:2014ott}
				R.~Aaij \textit{et al.} [LHCb],
				Phys. Rev. Lett. \textbf{113}, 162001 (2014)
				
				\bibitem{LHCb:2016lxy}
				R.~Aaij \textit{et al.} [LHCb],
				Phys. Rev. D \textbf{94}, no.7, 072001 (2016)
				
				\bibitem{FOCUS:2003gru}
				J.~M.~Link \textit{et al.} [FOCUS],
				Phys. Lett. B \textbf{586}, 11-20 (2004)
				
				\bibitem{LHCb:2020bls}
				R.~Aaij \textit{et al.} [LHCb],
				Phys. Rev. Lett. \textbf{125}, 242001 (2020)
				
				\bibitem{Wang:2023hpp}
				B.~Wang, K.~Chen, L.~Meng and S.~L.~Zhu,
				Phys. Rev. D \textbf{109}, no.3, 034027 (2024)
				
				\bibitem{LHCb:2021vvq}
				R.~Aaij \textit{et al.} [LHCb],
				Nature Phys. \textbf{18}, no.7, 751-754 (2022)
				
				\bibitem{Ali:2019roi}
				A.~Ali, L.~Maiani and A.~D.~Polosa,
				Cambridge University Press, 2019,
				ISBN 978-1-316-76146-5, 978-1-107-17158-9, 978-1-316-77419-9
				
				\bibitem{Chen:2020eyu}
				Y.~K.~Chen, J.~J.~Han, Q.~F.~L\"u, J.~P.~Wang and F.~S.~Yu,
				Eur. Phys. J. C \textbf{81}, no.1, 71 (2021)
				
				\bibitem{Wang:2020xyc}
				Z.~G.~Wang,
				Int. J. Mod. Phys. A \textbf{35}, no.30, 2050187 (2020)
				
				\bibitem{Agaev:2020mqq}
				S.~S.~Agaev, K.~Azizi, B.~Barsbay and H.~Sundu,
				Eur. Phys. J. A \textbf{57}, no.3, 106 (2021)
				
				\bibitem{Padmanath:2015era}
				M.~Padmanath, C.~B.~Lang and S.~Prelovsek,
				Phys. Rev. D \textbf{92}, no.3, 034501 (2015)
				
				\bibitem{Meng:2021jnw}
				L.~Meng, G.~J.~Wang, B.~Wang and S.~L.~Zhu,
				Phys. Rev. D \textbf{104}, no.5, 051502 (2021)
				
				\bibitem{Fleming:2021wmk}
				S.~Fleming, R.~Hodges and T.~Mehen,
				Phys. Rev. D \textbf{104}, no.11, 116010 (2021)
				
				\bibitem{Ortega:2016mms}
				P.~G.~Ortega, J.~Segovia, D.~R.~Entem and F.~Fernandez,
				Phys. Rev. D \textbf{94}, no.7, 074037 (2016)
				
				\bibitem{Esposito:2016noz}
				A.~Esposito, A.~Pilloni and A.~D.~Polosa,
				Phys. Rept. \textbf{668}, 1-97 (2017)
				
				\bibitem{Liu:2019zoy}
				Y.~R.~Liu, H.~X.~Chen, W.~Chen, X.~Liu and S.~L.~Zhu,
				Prog. Part. Nucl. Phys. \textbf{107}, 237-320 (2019)
				
				\bibitem{Lodha:2025ffp}
				C.~Lodha and A.~K.~Rai,
				[arXiv:2505.01195 [hep-ph]].
				
				\bibitem{Lodha:2024}
				C.~Lodha and A.~K.~Rai,
				\textit{Indian J. Phys.} (2024),
				
				%
				\bibitem{Lodha:2024bwn}
				C.~Lodha and A.~K.~Rai,
				Few Body Syst. \textbf{65}, no.4, 99 (2024)
				
				\bibitem{Lodha:2024qby}
				C.~Lodha and A.~K.~Rai,
				Eur. Phys. J. Plus \textbf{139}, no.7, 663 (2024)
				
				
				
				\bibitem{Lucha:1995zv}
				W.~Lucha and F.~F.~Schoberl,
				
				
				\bibitem{Koma:2006si}
				Y.~Koma, M.~Koma and H.~Wittig,
				Phys. Rev. Lett. \textbf{97}, 122003 (2006)
				
				
				\bibitem{Muta:2010xua}
				T.~Muta,
				``Foundations of Quantum Chromodynamics: An Introduction to Perturbative Methods in Gauge Theories, (3rd ed.),''
				World Scientific, 2010,
				ISBN 978-981-279-353-9
				
				
				
				\bibitem{Lucha:1991vn}
				W.~Lucha, F.~F.~Schoberl and D.~Gromes,
				Phys. Rept. \textbf{200}, 127-240 (1991)
				
				
				
				\bibitem{Voloshin:2007dx}
				M.~B.~Voloshin,
				Prog. Part. Nucl. Phys. \textbf{61}, 455-511 (2008)
				
				
				\bibitem{Debastiani:2017msn}
				V.~R.~Debastiani and F.~S.~Navarra,
				Chin. Phys. C \textbf{43}, no.1, 013105 (2019)
				
				\bibitem{Fredriksson:1981mh}
				K.~Fredriksson,
				\newblock \emph{A theoretical study of the meson spectrum in the Bethe-Salpeter formalism},
				\newblock Nucl. Phys. B \textbf{186}, 478-490 (1981).
				
				
				\bibitem{Debastiani:2016msc}
				V.~R.~Debastiani,
				\newblock \emph{Espectroscopia do Todo-Charme Tetraquark},
				\newblock Master's thesis, Instituto de Física, Universidade de São Paulo, 2016.
				
				
				\bibitem{Faustov:2021hjs}
				R.~N.~Faustov, V.~O.~Galkin and E.~M.~Savchenko,
				Universe \textbf{7}, no.4, 94 (2021)
				
				\bibitem{Chen:2023cws}
				J.~K.~Chen, X.~Feng and J.~Q.~Xie,
				JHEP \textbf{10}, 052 (2023)
				
				\bibitem{Yin:2021uom}
				P.~L.~Yin, Z.~F.~Cui, C.~D.~Roberts and J.~Segovia,
				Eur. Phys. J. C \textbf{81}, no.4, 327 (2021)
				
				%
				\bibitem{Yu:2006ty}
				Y.~M.~Yu, H.~W.~Ke, Y.~B.~Ding, X.~H.~Guo, H.~Y.~Jin, X.~Q.~Li, P.~N.~Shen and G.~L.~Wang,
				Commun. Theor. Phys. \textbf{46}, 1031-1039 (2006)
				
				
				\bibitem{Giannuzzi:2019esi}
				F.~Giannuzzi,
				Phys. Rev. D \textbf{99}, no.9, 094006 (2019)
				
				%
				\bibitem{Gutierrez-Guerrero:2021fuj}
				L.~X.~Gutierrez-Guerrero, J.~Alfaro and A.~Raya,
				Int. J. Mod. Phys. A \textbf{36}, no.24, 2150171 (2021)
				
				%
				\bibitem{Tiwari:2022kdu}
				R.~Tiwari, J.~Oudichhya and A.~K.~Rai,
				Int. J. Mod. Phys. A \textbf{38}, no.33n34, 2341007 (2023)
				
				%
				\bibitem{Kleiv:2013dta}
				R.~T.~Kleiv, T.~G.~Steele, A.~Zhang and I.~Blokland,
				Phys. Rev. D \textbf{87}, no.12, 125018 (2013)
				
				%
				\bibitem{deOliveira:2023hma}
				T.~de Oliveira, D.~Harnett, R.~Kleiv, A.~Palameta and T.~G.~Steele,
				Phys. Rev. D \textbf{108}, no.5, 054036 (2023)
				
				%
				\bibitem{Watanabe:2021nwe}
				K.~Watanabe,
				Phys. Rev. D \textbf{105}, no.7, 074510 (2022)
				
				%
				\bibitem{Bjorken:1988kk}
				J.~D.~Bjorken,
				Nucl. Phys. B Proc. Suppl. \textbf{11}, 325-341 (1989)
				
				
				\bibitem{Shifman:1991vm}
				M.~A.~Shifman,
				Nucl. Phys. B \textbf{388}, 346-362 (1992)
				
				%
				\bibitem{Faustov:2019mqr}
				R.~N.~Faustov, V.~O.~Galkin and X.~W.~Kang,
				Phys. Rev. D \textbf{101}, no.1, 013004 (2020)
				
				%
				\bibitem{Kamal:1995fr}
				A.~N.~Kamal, A.~B.~Santra, T.~Uppal and R.~C.~Verma,
				Phys. Rev. D \textbf{53}, 2506-2515 (1996)
				
				
				\bibitem{Zhang:2020dla}
				L.~Zhang, X.~W.~Kang, X.~H.~Guo, L.~Y.~Dai, T.~Luo and C.~Wang,
				JHEP \textbf{02}, 179 (2021)
				
				\bibitem{Manak}
				M. Parmar and A.K. Rai,
				Private communication
				
				
				%
				\bibitem{Yu:2022ngu}
				S.~Y.~Yu, X.~W.~Kang and V.~O.~Galkin,
				Front. Phys. (Beijing) \textbf{18}, no.6, 64301 (2023)
				
				%
				\bibitem{Biswas:2015aaa}
				A.~Biswas, N.~Sinha and G.~Abbas,
				Phys. Rev. D \textbf{92}, no.1, 014032 (2015)
				
				%
				\bibitem{LHCb:2013jjb}
				R.~Aaij \textit{et al.} [LHCb],
				JHEP \textbf{09}, 145 (2013)
				
				\bibitem{BaBar:2010zpy}
				P.~del Amo Sanchez \textit{et al.} [BaBar],
				Phys. Rev. D \textbf{82}, 111101 (2010)
				
				\bibitem{LHCb:2019juy}
				R.~Aaij \textit{et al.} [LHCb],
				Phys. Rev. D \textbf{101}, no.3, 032005 (2020)
				
				%
				\bibitem{LHCb:2015tsv}
				R.~Aaij \textit{et al.} [LHCb],
				Phys. Rev. D \textbf{92}, no.1, 012012 (2015)
				
				\bibitem{LHCb:2015klp}
				R.~Aaij \textit{et al.} [LHCb],
				Phys. Rev. D \textbf{92}, no.3, 032002 (2015)
				
				\bibitem{BaBar:2009pnd}
				B.~Aubert \textit{et al.} [BaBar],
				Phys. Rev. D \textbf{79}, 112004 (2009)
				
				\bibitem{Belle:2003nsh}
				K.~Abe \textit{et al.} [Belle],
				Phys. Rev. D \textbf{69}, 112002 (2004)
				
				\bibitem{BESIII:2019phe}
				M.~Ablikim \textit{et al.} [BESIII],
				Phys. Lett. B \textbf{804}, 135395 (2020)
				
				\bibitem{Belle:2004bvv}
				K.~Abe \textit{et al.} [Belle],
				Phys. Rev. Lett. \textbf{94}, 221805 (2005)
				
				\bibitem{LHCb:2015eqv}
				R.~Aaij \textit{et al.} [LHCb],
				Phys. Rev. D \textbf{91}, no.9, 092002 (2015)
				[erratum: Phys. Rev. D \textbf{93}, no.11, 119901 (2016)]
				
				
				%
				\bibitem{DELPHI:1998oyl}
				P.~Abreu \textit{et al.} [DELPHI],
				Phys. Lett. B \textbf{426}, 231-242 (1998)
				
				
				
			\end{thebibliography}
		\end{document}